\documentclass[journal]{IEEEtran}
\usepackage{comment}
\usepackage{cite}
\usepackage{hyperref}
\usepackage{url} 
\usepackage{amsmath,amssymb,amsfonts}
\usepackage{algorithmic}
\usepackage{graphicx}
\usepackage{textcomp}
\usepackage{booktabs}
\usepackage{soul}
\usepackage{xcolor}
\usepackage{xurl}
\usepackage{bm}
\usepackage{tabularx}
\usepackage{makecell}
\usepackage{tabularx,booktabs,makecell,ragged2e,array}

\definecolor{YInMnBlue}{HTML}{4E598C}
\definecolor{LapisBlue}{HTML}{255F85}
\definecolor{CelestialBlue}{HTML}{009DDC}
\definecolor{IndigoBlue}{HTML}{344966}
\definecolor{ResolutionBlue}{HTML}{14248A}

\usepackage{booktabs} 
\usepackage{xcolor}  
\usepackage{colortbl}

\usepackage{footmisc}

\ifCLASSINFOpdf
\else
\fi
\begin{document}

\title{
From Alignment to Advancement: Bootstrapping Audio-Language Alignment with Synthetic Data
}

\author{Chun-Yi Kuan and Hung-yi Lee~\thanks{
Chun-Yi Kuan and Hung-yi Lee are with the Graduate
Institute of Communication Engineering, National Taiwan University, 
Taipei 10617, Taiwan (e-mail: chunyi.kuan.tw@gmail.com; 
hungyilee@ntu.edu.tw).}}

\maketitle
\begin{abstract}
Audio-aware large language models (ALLMs) have recently made great strides in understanding and processing audio inputs. 
These models are typically adapted from text-based large language models (LLMs) through additional training on audio-related tasks.
This adaptation process presents two major limitations. 
First, ALLMs often suffer from catastrophic forgetting, where crucial textual capabilities like instruction-following are lost after training on audio data.
In some cases, models may even hallucinate sounds that are not present in the input audio, raising concerns about reliability.
Second, achieving cross-modal alignment between audio and language typically relies on large collections of task-specific question–answer pairs for instruction tuning, making it resource-intensive.
To address these issues, previous works have leveraged the backbone LLMs to synthesize general-purpose, caption-style alignment data. 
In this paper, we propose a data generation framework that produces contrastive-like training data, designed to enhance ALLMs’ ability to differentiate between present and absent sounds. 
We further extend our approach to multi-audio scenarios, enabling the model to either explain differences between audio inputs or produce unified captions that describe all inputs, thereby enhancing audio-language alignment. 
We refer to the entire ALLM training framework as bootstrapping audio-language alignment via synthetic data generation from backbone LLMs (BALSa).
Experimental results indicate that our method effectively mitigates audio hallucinations while reliably maintaining strong performance on audio understanding and reasoning benchmarks, as well as instruction-following skills. 
Moreover, incorporating multi-audio training further enhances the model's comprehension and reasoning capabilities. 
Overall, BALSa offers an efficient and scalable approach to developing ALLMs.
\end{abstract}

\begin{IEEEkeywords}
Audio-aware large language models, audio-language alignment
\end{IEEEkeywords}

%
\IEEEpeerreviewmaketitle

\section{Introduction}

The development of large language models (LLMs) has garnered increasing attention from the research community. Building on this foundation, researchers have explored integrating different modalities into LLMs beyond their strong textual capabilities. 
For instance, incorporating visual or audio modalities enables these models to perform visual or auditory understanding, thereby extending their capabilities beyond text processing. 
This multimodal integration allows LLMs to leverage their inherent linguistic proficiency alongside their understanding of other modalities to perform multimodal reasoning and question answering.

In the audio-language domain, audio-aware large language models~\cite{chu2023qwen, chu2024qwen2, tang2023salmonn, gong2023listen, gong2023joint, ghosh2024gama, kong2024audio, deshmukh2023pengi, kuan2024speech} (ALLMs) refer to models that integrate audio processing capabilities into LLMs, enabling them to understand and generate responses based on auditory inputs.
A critical step in developing these models is cross-modal alignment, where textual and auditory information are integrated to ensure effective reasoning across modalities. 

To achieve such alignment in practice, the dominant approach relies on instruction tuning with large-scale audio-language datasets~\cite{chu2023qwen, chu2024qwen2, tang2023salmonn, gong2023listen, gong2023joint}, typically composed of task-specific question-answer pairs. 
However, it is expensive to manually collect this kind of data, so some previous work~\cite{chu2023qwen, chu2024qwen2, gong2023listen, gong2023joint, tang2023salmonn} often turns to proprietary language models like ChatGPT~\cite{achiam2023gpt} to synthesize it using metadata from existing audio corpora.
Beyond the data curation effort, another fundamental issue with existing ALLMs is catastrophic forgetting.
Since most ALLMs are adapted from text-based LLMs, they inherently possess strong language understanding capabilities before being trained on audio tasks.
However, prior research has reported that currently developed ALLMs often lose their textual capabilities after being trained on audio data. 
This includes a decline in instruction-following behavior~\cite{key202502}, overfitting to pre-defined tasks~\cite{tang2023salmonn}, and limited generalization to unseen cross-modal reasoning tasks~\cite{tang2023salmonn}. 
In some cases, the models may even mistakenly identify sounds that are not present in the audio~\cite{kuan2024understanding}.

To address both the challenge of catastrophic forgetting and the burden of data curation, there is a growing trend toward leveraging a backbone LLM to generate synthetic data for audio-language alignment~\cite{fathullah2023towards, wang2023blsp, wang2024blsp, held2024distilling}.
Here, the backbone LLM refers to the large language model that serves as the initialisation of the core component of ALLM architectures.
Instead of relying on manually crafted or task-specific question-answer pairs, which are often generated using proprietary LLMs and complex rules, these approaches leverage the backbone LLM to synthesize training data in a prompt-driven, automatic, and extensible manner.
The developers only need to design simple generation prompts, and the backbone LLM can flexibly generate diverse data with minimal human effort.
While foundational datasets like Audio Dialogues~\cite{goel2024audio}, CoTA~\cite{xie2025audio}, and CompA~\cite{ghosh2023compa} have been instrumental, their creation methodologies differ fundamentally from an emerging paradigm: leveraging the backbone LLM itself to synthesize training data~\cite{fathullah2023towards, wang2023blsp, wang2024blsp, held2024distilling}. 
While this approach has been widely explored in the speech domain, it has not yet been extended to general audio.
This paper is the first to bridge this gap, demonstrating the advantages of this paradigm in simplicity, flexibility, and scalability for diverse general audio tasks.

This paper represents the framework, \textbf{BALSa}, short for \textbf{B}ootstrapping \textbf{A}udio-\textbf{L}anguage Alignment via \textbf{S}ynthetic Dat\textbf{a} Generation from Backbone LLMs.
Different from previous works~\cite{lu2024developing, fathullah2023towards, wang2023blsp, wang2024blsp, held2024distilling} using backbone LLMs for synthetic data generation, BALSa generates not only descriptive data, such as natural language descriptions of events occurring in the audio, but also contrastive data, which distinguish sounds that are present from those that are absent by simply varying the generation prompt\footnote{The incorporation of the synthetic negative samples was first proposed in our previous conference paper, \textbf{LISTEN} (\textbf{L}earning to \textbf{I}dentify \textbf{S}ounds \textbf{T}hrough \textbf{E}xtended \textbf{N}egative Samples)~\cite{kuan2025teaching}. This article extends that conference paper with a unified training framework, multi-audio modeling, and additional evaluations and experiments.}.
Like previous work of synthetic data generation, BALSa leverages the backbone LLM's understanding of audio metadata, such as sound event tags, to synthesize training data. Consequently, we only need the prompts like \textit{Repeat the audio} or \textit{List non-existent sounds in the audio} to efficiently generate diverse audio-language data tailored to various learning objectives.
BALSa offers a flexible and generalizable pipeline that facilitates audio-language alignment by learning from the behavioral patterns of the backbone LLM.
This highlights BALSa's potential as a general-purpose framework for synthesizing diverse alignment data for a wide range of audio understanding tasks.

In summary, this work introduces BALSa, a framework with several novel contributions that distinguish it from prior works.
First, we are the first to adapt the paradigm of backbone-LLM-driven synthetic data generation from the speech domain to general audio tasks, which cover a diverse range of non-speech events. 
This methodological shift allows for the generation of varied data formats tailored to specific learning objectives, such as description, comparison, or discrimination, simply by altering the generation prompts.
Second, we leverage this capability to address a critical challenge in the field: audio hallucination. 
Our proposed contrastive-style data generation, particularly the inclusion of synthetic negative samples, directly trains the model to differentiate present from absent sounds, effectively enhancing its reliability.
Finally, we demonstrate the scalability and efficiency of our framework. 
We extend it to multi-audio scenarios (BALSa-MA) to enable more complex reasoning, such as comparative analysis and joint captioning. 
Furthermore, our approach is highly data-efficient, achieving competitive results while using only a fraction of the training data required by previous state-of-the-art models.

To comprehensively evaluate model performance, we collect and construct multiple evaluation benchmarks, categorized into three major types: audio question answering, audio reasoning, and reliability and safety benchmarks. 
These benchmarks assess various aspects of the model’s capabilities, including basic audio understanding, advanced reasoning, and its reliability and safety in real-world applications.
Through these evaluations, we compare our proposed approach against popular baseline ALLMs in terms of audio-language alignment. 
Moreover, we investigate catastrophic forgetting and hallucination issues to further assess model robustness and reliability.
We provide illustrative examples on our demo page.~\footnote{\label{fn:demo}\url{https://kuan2jiu99.github.io/Balsa}}

In conclusion, our contributions are outlined as follows:
\begin{enumerate}
\item 
We present BALSa (Bootstrapping Audio-Language Alignment via Synthetic Data Generation from Backbone LLMs).
A critical feature of BALSa is a contrastive-style training strategy that incorporates synthetic negative samples, enabling models to distinguish clearly between present and absent sounds.
Experimental results demonstrate that BALSa effectively mitigates audio hallucinations, thereby enhancing the model's reliability in real-world applications.

\item 
Prior works~\cite{lu2024developing, fathullah2023towards, wang2023blsp, wang2024blsp, held2024distilling} have explored and validated the effectiveness of synthetic data generation using backbone LLMs. In these studies, the backbone LLM analyzes audio metadata and generates descriptions guided by designed prompts, but they focus primarily on the speech domain. 
This article is the first to apply this paradigm specifically to the general audio domain.
BALSa synthesizes audio-language alignment data using simple prompts from audio metadata, such as sound event tags, without requiring task-specific question-answer collections.

\item
We extend BALSa to support multi-audio processing by introducing BALSa-MA (Multi-Audio), which applies BALSa to scenarios involving multiple audio samples. 
In this approach, the backbone LLM both compares audio samples by generating descriptions of their differences and provides separate captions for each sample within the same response.
Experimental results indicate that exposing the model to multiple audio samples, whether through comparative explanations or joint captioning, enhances its audio understanding and improves its performance on evaluation benchmarks.

\item 
The BALSa framework achieves competitive performance with recent state-of-the-art models across reasoning, understanding, and trustworthiness benchmarks, while being data-efficient and even surpassing some models trained with substantially larger datasets.

\item
To comprehensively assess ALLMs, we conduct experiments on existing benchmarks, reproduce missing evaluation sets, and introduce new test sets to explore previously unexamined aspects.
Our evaluation covers three key areas: audio question answering, which assesses fundamental audio understanding; audio reasoning; and reliability and safety.
Through these evaluations, we systematically compare BALSa with mainstream models, examining the effectiveness of audio-language alignment and addressing challenges like audio hallucinations and instruction adherence.

\item 
We further perform complementary analyses to examine the robustness and generalizability of BALSa.
These include backbone replacement (Qwen3-4B vs. LLaMA-3.1-8B), LoRA-based fine-tuning, prompt design variations in synthetic data generation, and data scaling experiments.
The results show that BALSa remains stable under different backbones and training strategies, achieves more reliable performance when combining diverse prompts, and consistently benefits from larger training data sizes.
These findings highlight the framework’s robustness and potential for further improvements at scale.

\end{enumerate}

\section{Related Works}

There is already extensive research on audio and speech-related large language models. Due to space constraints, this paper discusses only the studies directly relevant to our work. For a comprehensive overview of the broader framework of audio and speech large language models (LLMs), readers are encouraged to consult the survey papers~\cite{arora2025landscapespokenlanguagemodels,ji2024wavchatsurveyspokendialogue,peng2025surveyspeechlargelanguage,cui2025recentadvancesspeechlanguage,wu2024audiolanguagemodeling,guo2025recentadvancesdiscretespeech,mousavi2025discreteaudiotokenssurvey,yang2025holisticevaluationlargeaudiolanguage}.

LLMs~\cite{touvron2023llama2, dubey2024llama, peng2023instruction, yang2024qwen2, zhao2023survey} can perform zero-shot learning through instruction tuning, which emerges when strong cross-modal alignment is achieved, particularly when a text-based LLM is integrated with multimodal encoders.
Following this trend, audio-aware large language models (ALLMs)~\cite{chu2023qwen, chu2024qwen2, tang2023salmonn, gong2023listen, gong2023joint, ghosh2024gama, kong2024audio, deshmukh2023pengi, deshmukh2025adiff, comanici2025gemini, xu2025qwen2, abouelenin2025phi, xie2025audio, goel2025audio, ghoshaudio, goel2024audio, ma2025audio, ding2025kimi, huang2025step, wu2025step} extend this paradigm to the audio domain by combining audio encoders with language models in a multi-stage training process.

Qwen-Audio~\cite{chu2023qwen} adopts a two-stage training strategy. 
In the first stage, multi-task pre-training is conducted using a manually curated dataset that includes tasks such as acoustic scene classification, sound event classification, and audio question answering. 
To unify the training format across tasks, Qwen-Audio introduces task-specific tags that indicate the nature of each task. 
This stage aims to build a foundational understanding of audio signals.
In the second stage, Qwen-Audio performs supervised instruction tuning to align the model more closely with human intent. 
To create training data, proprietary LLMs~\cite{achiam2023gpt, brown2020language, team2024gemini} are used to generate question–answer pairs based on text metadata, while additional audio-dialogue datasets are constructed using human annotations. 
Through this process, Qwen-Audio enhances the model’s ability in both audio understanding and reasoning.
Qwen2-Audio~\cite{chu2024qwen2} follows a similar approach, focusing on collecting high-quality supervised fine-tuning data through meticulous curation.
SALMONN~\cite{tang2023salmonn} also employs the same two-stage pipeline. 
During instruction tuning, it prioritizes tasks based on their relevance and manually collects task-specific datasets, including those for audio question answering. 
Like Qwen-Audio, it leverages proprietary LLMs to generate question–answer pairs grounded in the captions of existing audio corpora.
LTU~\cite{gong2023listen} constructs a large-scale audio question answering dataset, covering both closed-ended and open-ended formats. 
The closed-ended set includes tasks such as sound classification, audio captioning, and temporal analysis. Questions are generated using proprietary LLMs, while answers follow a fixed structure derived from rule-based algorithms. 
For open-ended data, which is more expensive and labor-intensive to create manually, LTU uses a rule-based pipeline known as audio instruction generation.
This method inputs metadata such as audio captions and sound event tags into carefully designed prompts, allowing proprietary LLMs to produce diverse and contextually appropriate audio QA pairs.

Recent models~\cite{ghosh2024gama, ghoshaudio, goel2025audio, xu2025qwen2, xie2025audio} further advance this multi-stage paradigm by strengthening the audio encoder, scaling data, or enhancing reasoning. 
Audio Flamingo 3~\cite{goel2025audio} introduces a unified AF-Whisper encoder and large curated datasets (e.g., AudioSkills-XL, AF-Think) with a curriculum strategy, enabling long-audio understanding, multi-audio dialogue, and on-demand chain-of-thought reasoning. 
Qwen2.5-Omni~\cite{xu2025qwen2} extends beyond audio to an end-to-end multimodal system, adopting a Thinker–Talker architecture for comprehension and real-time speech generation.
Audio-Reasoner~\cite{xie2025audio} builds on Qwen2-Audio but focuses on explicit reasoning, leveraging a 1.2M chain-of-thought dataset (CoTA) to yield large performance gains on reasoning benchmarks.

On the other hand, a distinct strategy for developing speech- or audio-aware LLMs focuses on leveraging the backbone LLM itself to generate synthetic data for alignment. 
These alignment data include general response-style texts conditioned on speech/audio, captions describing speech/audio under textual conditions, or continuation responses of speech/audio content based on textual prompts.
Several previous works~\cite{fathullah2023towards, lu2024developing, wang2023blsp, held2024distilling, wang2024blsp} have adopted this approach. 
Unlike studies that rely on externally generated task-specific question-answer pairs~\cite{chu2023qwen, chu2024qwen2, tang2023salmonn, gong2023listen, gong2023joint}, these works utilized a backbone language model to generate general-purpose alignment data directly, enabling a more streamlined and self-contained approach to integrating speech/audio with language models.
AudioChatLlama~\cite{fathullah2023towards} and DiVA~\cite{held2024distilling} used the backbone language model to generate responses conditioned on textual transcriptions.
BLSP~\cite{wang2023blsp} further extended this idea by generating continuations based on transcriptions, while BLSP-Emo~\cite{wang2024blsp} incorporated additional paralinguistic information like emotion to produce more expressive outputs.
DeSTA2~\cite{lu2024developing} further extended the use of paralinguistic cues to enhance the expressiveness of generated speech captions.

Inspired by these methods and motivated by the lack of research applying this strategy to general audio settings, our work investigates how to bootstrap audio–language alignment through synthetic data generation using a backbone LLM.
We further explore various extensions of this idea to evaluate its generality and effectiveness across different scenarios. 
For example, we exploit the flexibility and scalability of this framework to introduce negative samples and multi-audio scenario into the training process.
This article provides a practical way to effectively train ALLMs without the need to manually collect or rely on proprietary models to generate a large amount of task-specific question-answer data.

\section{Method}
\label{sec-method}

\subsection{Bootstrapping Audio-Language Alignment via Synthetic Data Generation from Backbone LLMs}
During the data construction phase, we aim to generate audio-language pairs while ensuring minimal textual discrepancies between the training data and the backbone LLM used in the ALLM. 
This design allows the model to maintain consistency in textual representation without modifying the backbone LLM’s parameters during subsequent training.

The rationale behind generating audio-language pairs using the backbone LLM is to enable the model to acquire audio understanding capabilities solely through an audio modality adapter. 
This approach ensures that the ALLM aligns its comprehension of audio with the text-based metadata information understood by the backbone LLM, without requiring adjustments to the backbone LLM itself.

\begin{figure*}[ht]
    \centering
    \includegraphics[width=0.90\textwidth]{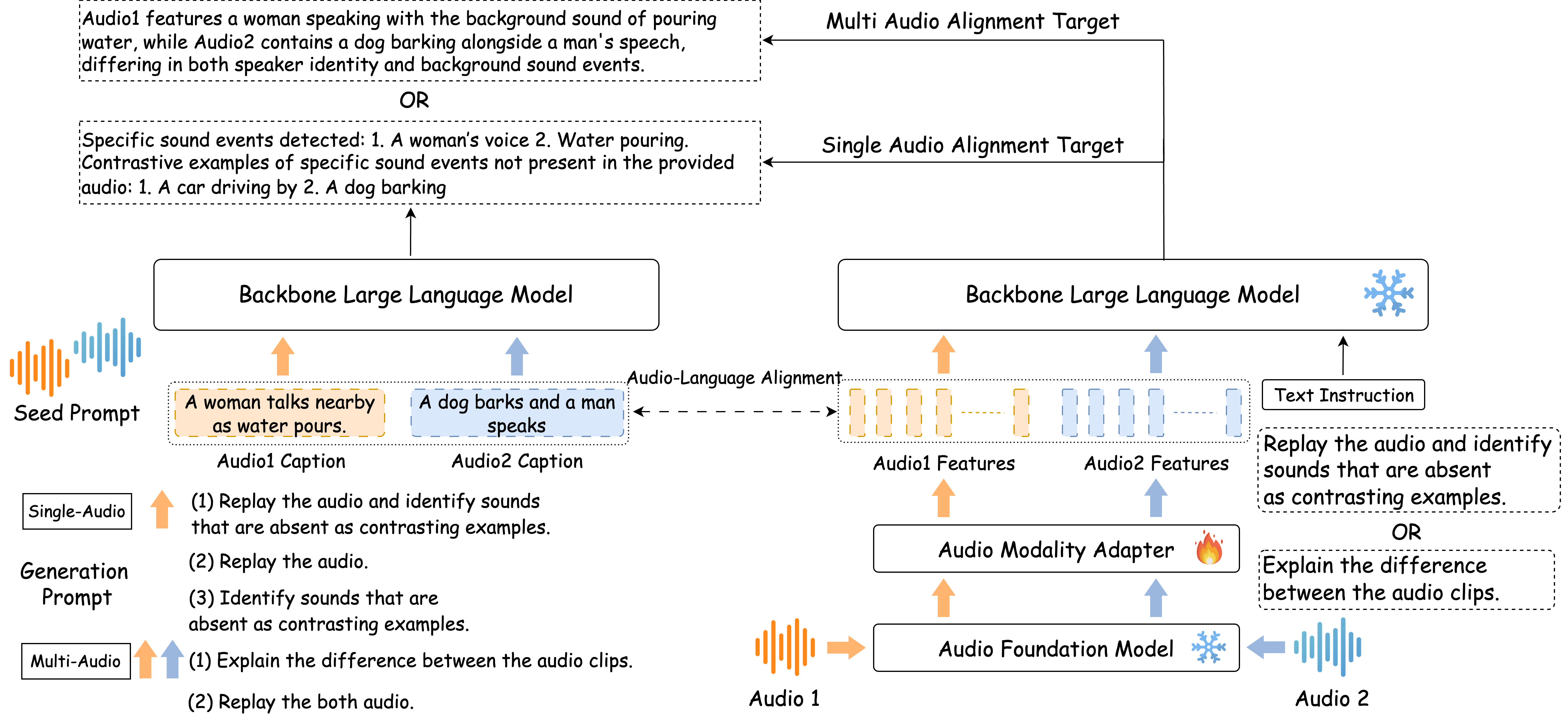} 
    \caption{
    The left diagram illustrates the Data Construction Stage, where a backbone LLM generates audio-language aligned descriptions for audio clips.
    The right diagram represents the Training Stage, where an audio modality adapter learns to align audio inputs while the backbone LLM remains frozen.
    In the single-audio scenario, the process follows only the orange arrows, whereas in the multi-audio scenario, both orange and blue arrows are involved, indicating the simultaneous alignment of multiple audio inputs.
    }
    \label{fig:overview-unified}
\end{figure*}

Next, we provide a detailed explanation of how LLMs are leveraged to generate audio-language alignment data.
While LLMs can not directly process audio, they excel at understanding and analyzing textual metadata associated with audio, such as sound events and audio captions.
To construct audio-language alignment data, we utilize the metadata from original audio datasets, which typically include two common types of annotations: human-annotated captions (${D}_{{caption}}$) and sound event tags (${D}_{{tag}}$).
To further minimize textual discrepancies and eliminate the need for manually designing task-specific question-answer pairs for formatting training data, we introduce a structured format called the seed prompt (${P}_{seed}$).
The seed prompt is derived from the original dataset's annotations and can take the form of either audio captions or sound event tags, depending on the available metadata.
We then append generation prompts (${P}_{gen}$) to the seed prompt and feed them into the LLM, prompting it to generate descriptions aligned with the audio content.
There are three types of generation prompts:

\noindent\textbf{Positive Samples Generation Prompt (${P}_{{pos}}$)} aims to generate descriptions of sound events that actually occur in the audio. 
For example, \textit{Can you describe what is in the audio}. 

\noindent\textbf{Negative Samples Generation Prompt (${P}_{{neg}}$)} aims to generate descriptions of sound events that are not present in the audio. 
For example, \textit{Identify sounds that are absent as contrasting examples}.

\noindent\textbf{Combined Samples Generation Prompt (${P}_{{comb}}$)} integrates both of the above prompts. 
It aims to generate descriptions of both the sound events that are present and those that are absent in the audio. 
For example, \textit{Replay the audio and identify sounds that are absent as contrasting examples}.

In summary, the final input prompt (${P}_{{final}}$) to the LLM can be expressed in Equation~\ref{equation-BALSa} and the overall pipeline is illustrated in Figure~\ref{fig:overview-unified}.
We intentionally adopt simple and direct prompts to establish a strong baseline, avoiding reliance on complex prompt engineering.

\begin{equation}
\begin{aligned}
P_{final} = & \, [\textit{Begin of audio}] \, P_{seed} \, [\textit{End of audio}] \, P_{gen} \\
& P_{seed} \in \{D_{caption}, D_{tag}\} \\
& P_{gen} \in \{P_{pos}, P_{neg}, P_{comb}\}.
\end{aligned}
\label{equation-BALSa}
\end{equation}

In Equation~\ref{equation-BALSa}, \textit{[Begin of audio]} and \textit{[End of audio]} are expressed in natural language format, with the purpose of indicating to the model that the information between them represents the content from the audio. 
When training audio-aware large language models, the position of ${P}_{{seed}}$ will be replaced with corresponding audio representation.
These generated responses to the input context are then used as our training targets.
Actual examples are demonstrated in Table~\ref{tab:actual_examples_BALSa}.
In this work, we refer to the whole data generation process in this subsection as \textbf{B}ootstrapping \textbf{A}udio-\textbf{L}anguage Alignment via \textbf{S}ynthetic Dat\textbf{a} Generation from Backbone LLMs (\textbf{BALSa}).

\begin{table}[t]
    \small
    \centering
    \caption{Examples produced under different generation prompts}
    \begin{tabular}{p{23em}}
    \toprule
    \textbf{Input Audio:} \\
    \textcolor{ResolutionBlue}{Audio Caption: A woman talks nearby as water pours.} \\
    \hline
    \textbf{Positive Samples Generation Prompt ($\boldsymbol{P}_{\boldsymbol{pos}}$):} \\
    Replay the audio. \\
    \textbf{Response:} \\
    Water pouring sounds, woman talking in the background. \\
    \hline
    \textbf{Negative Samples Generation Prompt ($\boldsymbol{P}_{\boldsymbol{neg}}$):} \\
    Identify sounds that are absent as contrasting examples. \\
    \textbf{Response:} \\
    Based on the provided audio, here are some specific sound events that are not present in the audio: \\
    1. A car driving by \hspace{0.05cm} 2. Birds chirping \hspace{0.05cm} 3. A dog barking \\
    \hline
    \textbf{Combined Samples Generation Prompt ($\boldsymbol{P}_{\boldsymbol{comb}}$):} \\
    Replay the audio and identify sounds that are absent as contrasting examples. \\
    \textbf{Response:} \\
    Specific sound events detected in the provided audio: \\
    1. A woman's voice \hspace{0.05cm} 2. Water pouring \\
    Contrastive examples of specific sound events not present in the provided audio: \\
    1. A car driving by \hspace{0.05cm} 2. A dog barking \hspace{0.05cm} 3. The door bang \\
    \bottomrule
\end{tabular}
\label{tab:actual_examples_BALSa}
\end{table}

\subsection{BALSa-MA: Bootstrapping Multi-Audio Learning via Difference Comparison and Joint Captioning}

BALSa can be extended to multi-audio scenarios, enabling the backbone LLM to both generate comparative explanations of differences between audio samples and independently caption each sample within the same response for training purposes.
The objective is to move beyond single-audio alignment by introducing a multi-audio learning framework that incorporates both discrimination-based comparison and parallel description-based captioning.
In this process, the model must first independently comprehend each audio sample before either distinguishing their differences or simultaneously generating separate captions for both audio samples within a unified response.
This dual-task formulation increases the learning complexity compared to single-audio alignment.
We refer to this multi-audio extended approach as BALSa-MA (Multi-Audio).

To implement this, we extend the generation prompt to accommodate multiple audio inputs. Here, we consider a two-audio scenario, denoted as first audio and second audio.
The seed prompt also can be either audio captions (${D}_{{caption}}$) or sound event tags (${D}_{{tag}}$), depending on the annotation format of the original dataset. 
We denote them as ${P}_{{seed, 1}}$ and ${P}_{{seed, 2}}$, respectively.
Next, we append the generation prompts (${P}_{gen}$) to both seed prompts and feed them into the LLM, enabling it to generate discrimination-based descriptions that compare the two audio samples.

In this study, we employ the following generation prompt:
\noindent\textbf{Discrimination Generation Prompt (${P}_{{diff}}$)}, which generates comparative descriptions highlighting the differences between two audio samples.
For instance, \textit{Explain the difference between the audio clips}.

\noindent\textbf{The Both Audio Captioning Prompt (${P}_{{cap, both}}$)}, which generates separate captions for each audio sample, providing distinct descriptions for both inputs within a single response.





In summary, the final input prompt (${P}_{{final}}$) to the LLM can be expressed in Equation~\ref{equation-BALSa-D} and the overall pipeline is illustrated in Figure~\ref{fig:overview-unified}.

\begin{equation}
\begin{aligned}
    P_{final,1} &= [\textit{Begin of audio1}] \, P_{seed,1} \, [\textit{End of audio1}] \\
    P_{final,2} &= [\textit{Begin of audio2}] \, P_{seed,2} \, [\textit{End of audio2}] \\
    P_{final}   &= P_{final,1} \, || \, P_{final,2} \, || \, P_{gen} \\
    P_{seed}    &\in \{D_{caption}, D_{tag}\} \\
    P_{gen}     &\in \{P_{diff}, P_{cap,both}\}.
\end{aligned}
\label{equation-BALSa-D}
\end{equation}

In Equation~\ref{equation-BALSa-D}, $\parallel$ denotes concatenation, while \textit{[Begin of audio]} and \textit{[End of audio]} are expressed in natural language format to serve two main purposes.
First, they indicate to the model that the enclosed content corresponds to the audio. 
Second, they provide flexibility for adaptation to multiple-audio formats.
When training audio-aware large language models, the positions of ${P}_{{seed, 1}}$ and ${P}_{{seed, 2}}$ will be replaced with the corresponding audio representations.
The generated responses based on the input context are then used as training targets.
Actual examples are presented in Table~\ref{tab:actual_examples_BALSa-D}.

\begin{table}[t]
    \small
    \centering
    \caption{Examples corresponding to different generation prompts.}
    \begin{tabular}{p{23em}}
    \toprule
    \textbf{Input Audio:} \\
    \textcolor{ResolutionBlue}{Audio1 Caption: A woman talks nearby while water pours.} \\
    \textcolor{ResolutionBlue}{Audio2 Caption: A dog barks and a man speaks.} \\
    \hline
    \textbf{Discrimination Generation Prompt ($\boldsymbol{P}_{\boldsymbol{pos}}$):} \\
    Explain the difference between the two audios. \\
    \textbf{Response:} \\
    Audio1 features a woman speaking with the background sound of pouring water, while Audio2 contains a dog barking alongside a man's speech, differing in both speaker identity and background sound events. \\
    \hline
    \textbf{Both Audio Captioning Prompt ($\boldsymbol{P}_{\boldsymbol{comb}}$):} \\
    Describe both audios. \\
    \textbf{Response:} \\
    Audio1: The sound of water pouring accompanies a woman speaking nearby. 
    Audio2: A man talks while a dog barks in the background. \\
    \bottomrule
\end{tabular}
\label{tab:actual_examples_BALSa-D}
\end{table}


\subsection{Model Design and Training Approach}

Figure~\ref{fig:overview-unified} illustrates the overall framework. The model comprises three primary components: an audio encoder, a backbone large language model, and an audio modality adapter.

\noindent\textbf{Audio Encoder}.
We employ \textit{Whisper}~\cite{radford2023robust}, a foundation model, as the audio encoder.
Previous studies~\cite{gong_whisperat} have demonstrated Whisper’s strong performance across various audio-related tasks, despite its original design being primarily for automatic speech recognition and speech translation.
Since Whisper follows an encoder-decoder architecture, we utilize only its encoder component. 
To preserve the advantages of the pre-trained model, the parameters of the audio encoder remain frozen.

\noindent\textbf{Backbone Large Language Model}.
This study adopts \textit{LLaMA-3.1-8B-Instruct}~\cite{dubey2024llama} and \textit{Qwen3-4B}, as the backbone large language models.
To maintain its original text-processing capabilities, we do not fine-tune any of its parameters.

\noindent\textbf{Audio Modality Adapter}.
The only trainable component is the audio modality adapter, which is randomly initialized.
This adapter projects the output representations extracted by the audio encoder into the input space of the backbone large language model.
Specifically, we employ a Qformer~\cite{li2023blip} architecture to extract audio features from the intermediate layers of the Whisper encoder.
These extracted features are aggregated using a weighted sum with learnable weights, followed by a linear projection layer that aligns the aggregated features with the input dimensions of the backbone large language model.
We optimize the entire architecture end-to-end with a next-token prediction loss, using the model’s own generated responses as learning targets.
To improve efficiency, we freeze the parameters of both the backbone LLM and the audio encoder.
Instead, we adopt a lightweight adapter-based training strategy~\cite{zhuminigpt, wang2023blsp, wang2024blsp, held2024distilling}, in which only the modality adapter is trained. 
This design not only conserves computational resources but also preserves the LLM’s original linguistic knowledge while enabling it to process audio through the adapter.

\section{Experimental Setup}

\subsection{Training Datasets}
The training datasets include AudioSet-20K~\cite{7952261}, AudioCaps~\cite{kim2019audiocaps}, FSD50K~\cite{fonseca2021fsd50k}, MACS~\cite{8682858}, ESC-50~\cite{piczak2015dataset}, UrbanSound8K~\cite{Salamon:UrbanSound:ACMMM:14}, Clotho~\cite{lipping2022clotho}, and VocalSound~\cite{gong_vocalsound}.
Among these datasets, AudioCaps, Clotho, and MACS provide ground-truth labels in the form of human-annotated audio captions, while the rest use sound event tags.
Table~\ref{tab:dataset_summary} summarizes the total duration of the training datasets used in our work and compares it with the training data durations reported in previous studies~\cite{chu2023qwen, chu2024qwen2, gong2023listen, gong2023joint, tang2023salmonn}.
The ``Duration'' column represents the original total duration of unique audio in each dataset, measured in hours and excluding any duplicate segments. 
The ``Equi. Duration'' column accounts for the effect of generating multiple aligned text descriptions per audio clip, leading to an increased effective training duration, also measured in hours. 
The number of samples represents the total number of data instances used during training.

\begin{table}[ht]
\scriptsize
\centering
\caption{
Summary of training datasets.
}
\begin{tabular}{lcccc}
\hline
\textbf{Dataset}          
& \textbf{Duration} 
& \textbf{Equi. Duration}
& \textbf{\# Samples (k)} \\ 
\hline
AudioSet-20K~\cite{7952261} 
& 60 & 121 & 42
\\
AudioCaps~\cite{kim2019audiocaps} 
& 138 & 545 & 145 
\\
Clotho~\cite{lipping2022clotho} 
& 18 & 196 & 40 
\\
MACS~\cite{8682858} 
& 11 & 22 & 33
\\
FSD50K~\cite{fonseca2021fsd50k} 
& 56 & 188 & 75 
\\
ESC50~\cite{piczak2015dataset} 
& 3 & 34 & 14 
\\
UrbanSound8K~\cite{Salamon:UrbanSound:ACMMM:14} 
& 10 & 19 & 17
\\
VocalSound~\cite{gong_vocalsound} 
& 20 & 40 & 34
\\
\hline
\textbf{Total (Ours)} 
& 316 & 1,164 & 400 \\
\hline
\textbf{Qwen2-Audio}~\cite{chu2023qwen,chu2024qwen2} 
& - & 10,000 & - 
\\
\textbf{SALMONN}~\cite{tang2023salmonn} 
& - & 1,044 & 370 
\\
\textbf{LTU}~\cite{gong2023listen}, \textbf{LTU-AS}\cite{gong2023joint} 
& - & 1,784 & 5,682
\\
\textbf{GAMA}~\cite{ghosh2024gama}
& - & - & 6,319
\\
\textbf{Qwen-2.5-Omni}~\cite{xu2025qwen2}
& - & 3,333,333 & -
\\
\textbf{Audio Flamingo 2}~\cite{ghoshaudio}
& - & 33,153 & 1,460
\\
\textbf{Audio Flamingo 3}~\cite{goel2025audio}
& - & 54,409 & 2,600
\\
\textbf{Audio-Reasoner}~\cite{xie2025audio}
& - & - & 1,200
\\
\textbf{Phi-4-Multimodal-IT}~\cite{abouelenin2025phi}
& - & - & 17,000
\\
\hline
\end{tabular}
\label{tab:dataset_summary}
\end{table}

\subsection{Model Selection and Training Setups}

Our framework integrates two key foundation models: LLaMA-3.1-8B-Instruct\footnote{\scriptsize{\url{huggingface.co/meta-llama/Llama-3.1-8B-Instruct}}}\cite{dubey2024llama} and the compact Whisper-small\footnote{\scriptsize{\url{huggingface.co/openai/whisper-small}}}\cite{radford2023robust}, which contains about 240 million parameters.
In addition, to examine the generalizability of our approach across different language model backbones, we further trained a set of models by replacing Llama-3.1-8B-Instruct with Qwen3-4B~\footnote{\scriptsize{\url{huggingface.co/Qwen/Qwen3-4B}}}~\cite{yang2025qwen3}, using several representative training data configurations.
To connect these components, followed by~\cite{zhuminigpt, wang2023blsp, wang2024blsp, held2024distilling}, we developed a specialized modality adapter, featuring a two-layer transformer decoder architecture. 
This adapter uses 64 specialized vectors to capture and process audio features from the encoder's hidden states.
In our implementation strategy, we maintain the original architecture of both LLaMA (or Qwen3-4B in the additional experiments) and Whisper unchanged. 
Instead, we focused our training efforts solely on the modality adapter, which introduces approximately 22 million trainable parameters. 
The system processes information by first converting text instructions through LLaMA's embedding layer, then combining them with the processed audio features.

To enhance the model’s ability to handle multi-audio inputs, a progressive learning strategy is adopted, where the model is first trained on single-audio tasks before transitioning to multi-audio scenarios, following a curriculum learning approach.
For the technical implementation, we built our system using the Hugging Face Transformers\cite{wolf2020huggingfacestransformersstateoftheartnatural} framework.
Following our progressive learning strategy, the training process consisted of two stages: first, we pretrained the model on single-audio tasks for five epochs to establish a strong baseline; 
second, we fine-tuned it under both positive and negative data settings on multi-audio data for two epochs to enhance its ability to process multiple inputs.
We used the Adam optimization algorithm with a cosine annealing learning schedule and 2,000 warm-up steps.
All experiments were conducted on two NVIDIA RTX 6000 Ada Generation GPUs, with a global batch size of 16 and a learning rate of 1e-4.

\subsection{Baseline Models and Evaluation Setups}

In this study, we incorporate widely used, well-documented, and fully open-source models as baselines for comparison. 
These include Qwen-Audio-Chat~\cite{chu2023qwen}, Qwen2-Audio-Instruct~\cite{chu2024qwen2}, Qwen2.5-Omni-Instruct~\cite{xu2025qwen2}, SALMONN~\cite{tang2023salmonn} (including 7B and 13B versions), LTU~\cite{gong2023listen}, LTU-AS~\cite{gong2023joint}, GAMA~\cite{ghosh2024gama}, Audio Flamingo 2~\cite{ghoshaudio}, Audio Flamingo 3~\cite{goel2025audio}, Audio-Reasoner~\cite{xie2025audio}, and Phi-4-Multimodal-Instruct~\cite{abouelenin2025phi}, as well as proprietary commercial models like Gemini-Pro-1.5~\cite{team2024gemini}, Gemini-Flash-2.5~\cite{comanici2025gemini}, Gemini-Pro-2.5~\cite{comanici2025gemini}, and GPT-4o-Audio~\cite{achiam2023gpt}.
Since most ALLMs are instruction-tuned for open-ended responses~\cite{chu2023qwen, chu2024qwen2, tang2023salmonn, gong2023listen, gong2023joint}, they generate answers based on user queries rather than selecting from predefined choices. 
To align with this design, we extract relevant answers from model outputs using a structured evaluation approach. 
All experiments adopt a consistent greedy decoding strategy with a maximum output length of 512.

To facilitate the extraction of relevant answers during evaluation, we implement robust regular expressions to extract key information from model responses.
The extracted content is then matched to the provided options via string-based comparison. 
To minimize potential biases, the answer options in the original question are randomized before evaluation. 
To ensure concise questions, we randomly select three incorrect options along with the correct answer and then shuffle their order in the multiple-choice format.

Once the relevant answers are extracted and mapped to the given options, we proceed with the evaluation process using standardized metrics. 
If a model's response cannot be parsed using the aforementioned approach, it is considered incorrect. 
Our evaluation strategy is structured as follows. 
We report the weighted F1 score for multi-class classification tasks. 
For other tasks with their own evaluation protocols, we follow established methodologies~\cite{kuan2024understanding, key202502, yang2025sakura, sakshi2024mmau, huang2024dynamic, huang2024dynamic2}.
In our experiment results, bold text indicates the best performance, while underlined text represents the second-best.
We further detail the evaluation methods for each benchmark.

\noindent\textbf{Audio Question Answering Benchmark}. 
Accuracy is adopted as the primary metric for ClothoAQA~\cite{lipping2022clotho}, which comprises both binary and non-binary classification tasks.

\noindent\textbf{Audio Reasoning Benchmark}. 
We report weighted F1 score for multi-class classification tasks. 
For Synonym-Hypernym Test, we also report weighted precision and recall.
In addition, the original MMAU~\cite{sakshi2024mmau} benchmark uses micro-averaged accuracy as its evaluation metric.
The subsequent MMAR~\cite{ma2025mmar} benchmark focuses on more challenging deep reasoning audio tasks, while adopting a similar evaluation approach to MMAU.
The SAKURA~\cite{yang2025sakura} test set follows a multiple-choice format, and we evaluate model performance using three key metrics. 
We report overall accuracy across all questions, as well as separate accuracy scores for single-hop and multi-hop questions. 

\noindent\textbf{Audio Hallucination Benchmark}. 
Following previous studies~\cite{kuan2024understanding}, we compute overall accuracy and F1 scores for questions where the correct answer is ``Yes'' or ``No''. 
As an additional reference, we also report the proportion of cases where the model responds with ``Yes''.

\noindent\textbf{Instruction-Following Benchmark}. 
We report the accuracy of each instruction-following task, the overall average accuracy, and the performance gap between the evaluated model and its backbone LLM.

\subsection{Ablation Study}

First, in the single-audio training scenario, to ensure a fair evaluation of contrastive-like methods using BALSa, we conducted the following experiments to investigate the role of negative samples.

\noindent\textbf{Positive-only Training Data:} 
The training set contains only positive samples, with a total of $2N$ data points.
    
\noindent\textbf{Positive and Negative Training Data:} 
The training set includes both positive and negative samples, each with $N$ data points, resulting in a total of $2N$ data points.
    
\noindent\textbf{Combined Training Data:} 
The training set consists of combined samples, as described in Section~\ref{sec-method}. 
A combined sample includes both sound events that are present and those that are absent within a single sample. 
In contrast, positive samples only contain present sound events, while negative samples only contain absent sound events. 
Under the combined setup, there are $N$ data points. 
Since each data point carries information about both present and absent sound events, it is equivalent to having a total of $2N$ data points.

Second, to validate that data generated by the backbone LLM outperforms data generated by an external LLM under this setting, we also examine a scenario where a stronger LLM is used to generate synthetic data via our proposed pipeline, BALSa.
Specifically, we select Gemini-1.5-Pro to synthesize the positive-only, positive and negative, and combined training data.
The same methodology is applied, but a more advanced model is used to generate the data.

Third, we compare two approaches for generating training data for audio-language alignment: our method, which allows the backbone LLM to freely generate captions based on meta-information, and another that provides it with collected questions and asks it to generate corresponding answers.
To investigate this, we leverage the questions (${Q}_{{OpenAQA}}$) from the OpenAQA dataset~\cite{gong2023joint} and apply the BALSa pipeline to generate corresponding answers using the backbone LLM. 
Specifically, OpenAQA contains 5 million instances, from which we select a subset that aligns with the audio data used in this work. 
As a result, the final training dataset also consists of $2N$ data points.
In summary, the final input prompt (${P}_{{final}}$) to the backbone LLM is formulated in Equation~\ref{equation-BALSa-OpenAQA}. This setting refers to BALSa-OpenAQA.

\begin{equation}
\begin{aligned}
P_{final} = & \, [\textit{Begin of audio}] \, P_{seed} \, [\textit{End of audio}] \, P_{gen} \\
& P_{seed} \in \{D_{caption}, D_{tag}\} \\
& P_{gen} \in \{Q_{OpenAQA}\}.
\end{aligned}
\label{equation-BALSa-OpenAQA}
\end{equation}
For comparison, we also directly use question-answer pairs derived from the OpenAQA dataset as our training data.

Fourth, we explore an alternative approach where, instead of using BALSa to generate captioning data, we employ pre-designed description-based prompts like \textit{"Describe the audio."}, which is the rule-based templates. 
In this setting, the training targets are the corresponding ground truth captions or sound event tags of the audio.
The total training dataset also consists of $2N$ data points.

Beyond the single-audio scenario, we conduct ablation experiments on the multi-audio scenario, BALSa-MA, an extended version of BALSa. 
In BALSa-MA, we aim to enable models to learn from multiple audio sources, including explaining the differences between two audio samples and captioning both simultaneously. 
For conciseness, in our experiments, we adapt BALSa-MA to a two-audio-sample setting.

To ensure a fair evaluation of different methods for audio interpretation, we design the following experiments.

\noindent\textbf{Discrimination-only Training Data:}
The training set contains only discrimination samples, with a total of $2M$ data points.

\noindent\textbf{Captioning-only Training Data:}
The training set contains only captioning samples, with a total of $2M$ data points.

\noindent\textbf{Joint Discrimination-Captioning Training Data:}
This training set is a balanced combination of discrimination and captioning samples, with each type comprising 50\% of the dataset, totaling $2M$ data points.

In addition to the primary experiments, we conducted four complementary studies to evaluate different aspects of our framework:
(1) replacing the LLaMA-3.1-8B backbone with Qwen3-4B to assess backbone generalizability;
(2) applying LoRA fine-tuning instead of keeping the backbone frozen to study the effect of fine-tuning strategy;
(3) varying prompts used in synthetic data generation to examine the impact of prompt design on model performance; and
(4) data scaling analysis, where we evaluate performance under different proportions of the training set (10\%, 25\%, 50\%, 75\%, and 100\%) to analyze the scalability of the BALSa approach.

\section{Evaluation Benchmarks}

In this work, we categorize evaluation benchmarks into three types: audio question answering benchmark, audio reasoning benchmark, and reliability and safety benchmark.

\subsection{Audio Question Answering Benchmark}
This benchmark focuses on evaluating a model’s fundamental understanding of audio.
It measures abilities like sound event recognition, acoustic scene classification, and temporal comprehension.
By designing diverse instruction-following questions (e.g., Select the correct category for the sound: Wind, Machine, Rain, Fly.), the benchmark tests whether a model can accurately interpret different aspects of an audio signal and provide reasonable, correct answers based on the given auditory information.
For this benchmark, we have collected and organized the following evaluation datasets.

\noindent\textbf{Clotho-AQA}: 
Clotho-AQA~\cite{lipping2022clotho} is a widely used dataset designed for audio question answering (AQA) tasks, where models generate natural language answers based on audio and corresponding questions. 

\noindent\textbf{Nonspeech7k AQA}:
Nonspeech7k~\cite{nonspeech} is a dataset of human non-speech sounds collected from real-life environments, covering seven sound categories: breath, cough, cry, laugh, scream, sneeze, and yawn. 
We utilize all 725 audio instances from the test split of the Nonspeech7k dataset for evaluation.

\noindent\textbf{CochlScene AQA}:
CochlScene~\cite{jeong2022cochlscene} is a crowd-sourced dataset designed for acoustic scene classification, covering 13 different acoustic scenes. 
The primary goal of this task is to determine the environment in which a given audio recording was captured based on its acoustic characteristics. 
For evaluation, we sample 100 audio clips from each category in the test split, resulting in a total of 1,300 test instances.

\noindent\textbf{EDANSA-2019 AQA}:
The Ecoacoustic Dataset from Arctic North Slope Alaska (EDANSA-2019)~\cite{ccoban2022edansa} contains recordings collected from the North Slope of Alaska in the summer of 2019.
As an Arctic dataset, it represents an out-of-distribution domain rarely covered by public training corpora, posing a challenge for models to generalize to unfamiliar acoustic environments.
For evaluation, we used sound events with sufficient samples: fly, bird, woof, wind, rain, aircraft, and machine.
Due to limited data, fly (62 instances) and woof (15 instances) have fewer samples, while the remaining categories each include 200 randomly selected clips, totaling 1,077 instances.

\subsection{Audio Reasoning Benchmark}
This benchmark focuses on evaluating a model’s advanced reasoning capabilities in audio understanding. 
Models must go beyond recognition to reason and infer from audio.
For this benchmark, we utilize the following datasets.

\noindent\textbf{Synonym and Hypernym Test}:
A previous study~\cite{ccoban2024mllms} introduced this test to examine whether models understand audio and its semantic ties to text. Synonyms capture similar meanings, while hypernyms capture broader categories. We leverage these relationships to construct prompts that probe similarity and hierarchy in audio reasoning. For example, given an audio of a songbird chirping, one question is: \textit{Is the sound from an object that is a type of chordate?} (Answer: yes). This benchmark tests whether models truly capture semantic relationships in audio, rather than relying on surface statistical associations. Following the original setup with AudioCaps, we derive 636 evaluation instances and report weighted precision, recall, and F1 for yes/no questions.

\noindent\textbf{Audio Entailment}:
A previous study~\cite{deshmukh2024audio} introduced audio entailment to assess a model’s deductive reasoning: determining whether a textual hypothesis logically follows from an audio premise. Conclusions are labeled as entailment (audio affirms the hypothesis), contradiction (audio disproves it), or neutral (insufficient evidence). 
We frame this as a multiple-choice task where the model selects the correct conclusion. 
For example, given an audio clip, the prompt may ask: \textit{Which hypothesis is entailed? (A) A quiet room with no movement or wind. (B) A motorcycle is overtaking cars on a windy road. (C) Vehicles are moving on a road with wind noise present.} Answer order is randomized. The dataset combines Clotho~\cite{lipping2022clotho} and AudioCaps~\cite{kim2019audiocaps}, and performance is measured by overall accuracy.

\noindent\textbf{SAKURA}:
The SAKURA benchmark~\cite{yang2025sakura} evaluates multi-hop reasoning in audio understanding, and we adopt its audio-track test set. 
Each audio clip is paired with two types of questions: single-hop and multi-hop. 
Single-hop questions ask directly about sound attributes, such as \textit{Out of the animals listed below, which most closely matches the audio?} Multi-hop questions require combining audio understanding with text-based inference, e.g., \textit{Considering the sound and the natural tendencies of the animal, which behavior is most representative? (a) Wagging its tail (b) Jumping between branches (c) Hiding in shells (d) Flapping its wings}.

\noindent\textbf{MMAU}:
We use the official MMAU-v05.15.25 benchmark~\cite{sakshi2024mmau}, a massive multi-task audio understanding and reasoning benchmark designed to evaluate multimodal audio understanding models on tasks requiring expert-level knowledge and complex reasoning.
In this study, we focus on assessing model performance specifically along the audio dimension, using the test-mini split for evaluation.
The evaluation settings follow the original benchmark’s guidelines.~\footnote{\url{github.com/Sakshi113/MMAU}}

\noindent\textbf{MMAR}:
A large-scale multi-task audio reasoning benchmark~\cite{ma2025mmar} covering diverse real-world sound, music, and speech scenarios, collected from internet videos and refined through rigorous quality checks.
It hierarchically categorizes questions into four reasoning layers (Signal, Perception, Semantic, Cultural), each requiring multi-step reasoning and, in some cases, domain-specific expertise.
Evaluation follows the original benchmark setup.\footnote{\url{github.com/ddlBoJack/MMAR}}

\subsection{Reliability and Safety Benchmark}
The Reliable and Safety Benchmark focuses on evaluating the trustworthiness and safety of ALLMs. 
We select the following evaluation benchmark.

\noindent\textbf{Audio Hallucination}:
Our previous work~\cite{kuan2024understanding} introduced benchmarks for object hallucination in audio, where models incorrectly infer non-existent sound events. 
Following the methodology of POPE~\cite{li2023evaluating}, we cast hallucination detection as a binary task (Yes/No). 
To test robustness, we design four prompt variants (e.g., Is there a sound of [object]?, Does the audio contain...?). 
The benchmark is built from AudioCaps, sampling 407 clips with over three sound events; ground-truth objects are treated as positives, while three randomly chosen absent objects per clip serve as negatives. 
We report precision, recall, and F1 separately for yes and no cases.
    
\noindent\textbf{Instruction Following Ability Evaluation}:
Prior work~\cite{key202502} shows that most ALLMs struggle with instruction following, exhibiting greater forgetting than text-only LLMs. 
We evaluate this ability using both close-ended (strict formatting) and open-ended (flexible, creative) questions. Performance is measured by instruction-following rate (IFrate) and forgetting rate ($\Delta$), where smaller $\Delta$ indicates less forgetting. 
Note that since we cannot directly access proprietary models like Gemini, we are unable to calculate the forgetting rate. 
Similarly, for the Qwen-Audio models~\cite{chu2023qwen, chu2024qwen2}, which are finetuned from the pretrained Qwen-Audio base model~\cite{chu2023qwen}, it is not feasible to compute the IFrate difference between ALLMs and their text-only LLM.

\section{Experimental Results}

\subsection{Performance of BALSa}
\begin{table*}[ht]
\scriptsize
\centering
\caption{
Evaluation results for the audio question answering and audio hallucination benchmark.
F1 (Y) and F1 (N) represent the F1 scores for cases where the correct answer is yes and no, respectively. 
F1 (W) denotes the weighted F1 score. 
Acc refers to accuracy, and Yes indicates the percentage of responses where the model answers yes. 
(Unit: \%)
}
\begin{tabular}{l c c c c cccc cc}
\toprule
\textbf{} 
& \multicolumn{1}{c}{\textbf{EDANSA}} 
& \multicolumn{1}{c}{\textbf{ClothoAQA}} 
& \multicolumn{1}{c}{\textbf{CochlScene}} 
& \multicolumn{1}{c}{\textbf{NonSpeech}}
& \multicolumn{4}{c}{\textbf{Audio Hallucination}}
& \multicolumn{2}{c}{\textbf{Audio Entailment}} \\
\cmidrule(r){2-2} \cmidrule(r){3-3} \cmidrule(r){4-4} \cmidrule(r){5-5} \cmidrule(r){6-9} \cmidrule(r){10-11}
\textbf{Models} 
& \textbf{F1} 
& \textbf{Acc} 
& \textbf{F1} 
& \textbf{F1}
& \textbf{F1 (Y)}
& \textbf{F1 (N)}
& \textbf{F1 (W)}
& \textbf{Yes}
& \textbf{Acc (C)}
& \textbf{Acc (A)}
\\
\midrule
Qwen-Audio-Chat\cite{chu2023qwen}
& 34.50
& 74.90
& 55.58
& 34.73
& 75.25 & 52.97 & 64.11 & 81.04
& 36.65 & 34.04
\\
Qwen2-Audio-Instruct\cite{chu2024qwen2} 
& 41.40
& 75.55
& 60.16
& 54.84
& 72.89 & 54.44 & 63.67 & 75.39 
& 42.97 & 46.19
\\
Qwen2.5-Omni-7B
& \textbf{70.81}
& 86.90
& \textbf{81.24}
& 54.03
& \textbf{89.72} & \underline{85.81} & \textbf{87.77} & 49.59
& \underline{56.81} & \textbf{62.81}
\\
SALMONN-7B\cite{tang2023salmonn}
& 26.36
& 76.56
& 37.99
& {78.29}
& 70.06 & 26.48 & 48.27 & 92.14
& 33.24 & 34.10
\\
SALMONN-13B\cite{tang2023salmonn}
& 29.96
& 83.99
& 31.06
& 77.24
& 76.94 & 60.07 & 68.50 & 76.78
& 35.25 & 30.81
\\
LTU\cite{gong2023listen} 
& 38.71
& 67.10
& 14.91
& 5.79
& 78.27 & 73.65 & 75.96 & 59.97
& 18.88 & 14.42
\\
LTU-AS\cite{gong2023joint} 
& 23.16
& 56.77
& 20.30
& 8.06
& 51.68 & 45.49 & 48.59 & 56.02 
& 26.95 & 28.64
\\
GAMA~\cite{ghosh2024gama}
& 52.13
& 72.34
& 65.66
& 38.81
& 81.42 & 77.41 & 79.41 & 59.75
& 29.35 & 25.87
\\
Audio Flamingo 2~\cite{ghoshaudio}
& 28.65
& 83.41
& 38.45
& 10.62
& 68.35 & 20.43 & 44.39 & 93.08
& 58.41 & 55.31
\\
Audio Flamingo 3~\cite{goel2025audio}
& 62.49
& \textbf{92.43}
& \underline{80.28}
& 68.55
& 69.53 & 78.42 & 73.97 & 32.92
& \textbf{59.49} & \underline{57.80}
\\
Audio-Reasoner~\cite{xie2025audio}
& 57.95 
& 70.89 
& 61.18 
& \underline{83.75}
& 71.61 & 73.51 & 72.59 & 36.12
& 10.02 & 23.54
\\
Phi-4-Multimodal-Instruct~\cite{abouelenin2025phi}
& 32.20
& 66.38
& 31.49
& 9.66
& 70.53 & 33.38 & 51.95 & 88.66
& 50.08 & 49.38
\\
Gemini-1.5-Pro\cite{team2024gemini} 
& 17.46
& 68.43
& 36.18
& 47.01
& 65.86 & 70.95 & 68.41 & 42.02
& 23.32 & 22.87
\\
Gemini-2.5-Flash~\cite{comanici2025gemini}
& 48.31 
& 71.62
& 68.80
& 65.64
& 83.08 & 81.21 & 82.14 & 49.75
& 17.94 & 16.27
\\
Gemini-2.5-Pro~\cite{comanici2025gemini}
& \underline{62.79}
& 82.18
& 57.89
& 77.60
& \underline{86.35} & 68.81 & 77.58 & 44.90
& 17.84 & 18.27
\\
GPT-4o-Audio~\cite{achiam2023gpt}
& 64.47
& \underline{87.36}
& 75.73
& 74.07
& 84.62 & 50.71 & 67.67 & 51.8
& 18.50 & 22.80
\\
\midrule
BALSa (Ours) \\
-- Llama-3.1, only-positive
& 45.57
& 75.11
& 39.28
& 54.17
& 73.05 & 55.15 & 64.10 & 74.81
& 50.88 & 52.41
\\
-- Llama-3.1, positive-negative
& 42.06
& 74.96
& 42.87
& 53.66
& 80.18 & 74.04 & 77.1 & 63.69
& 56.30 & 57.16
\\
-- Llama-3.1, combined
& 48.56
& 80.93
& 42.48
& 52.17
& 77.21 & 71.23 & 74.22 & 61.58
& 46.92 & 49.00
\\
-- Qwen3, only-positive
& 41.26 
& 80.79
& 51.97
& 48.47
& 82.44 & 80.44 & 81.44 & 55.41
& 42.97 & 48.80
\\
\midrule
BALSa-MA (Ours)
\\
-- Llama-3.1, discrimination
& 50.97
& 77.87 
& 44.77
& 68.65
& 85.35 & 84.66 & 85.00 & 52.29
& 54.67 & 55.66
\\
-- Llama-3.1, captioning
& 53.80
& 78.75
& 44.47
& 70.03
& 84.00 & 84.93 & 84.47 & 47.01
& 56.40 & 56.59
\\
-- Qwen3, discrimination
& 51.54
& 85.74
& 58.35
& \textbf{85.02}
& 85.87 & \textbf{86.82} & \underline{86.34} & 46.52
& 42.87 & 54.74
\\
\bottomrule
\end{tabular}
\label{tab:evaluation-results-part1}
\end{table*}
In Table~\ref{tab:evaluation-results-part1}, our proposed models trained using the BALSa method achieve competitive performance against a wide range of strong baseline models, especially when considering their training efficiency. 
For instance, in ClothoAQA, while models like Audio-Flamingo3 and Qwen-2.5-Omni set the state-of-the-art, our models perform robustly and are on par with other strong baselines like Qwen2-Audio.
In benchmarks like NonSpeech and CochlScene AQA, although top-performing proprietary models like Gemini-2.5-Pro and models like Audio-Flamingo 3 lead, our models still attain solid results.

\begin{figure}
    \centering
    \includegraphics[width=0.45\textwidth]{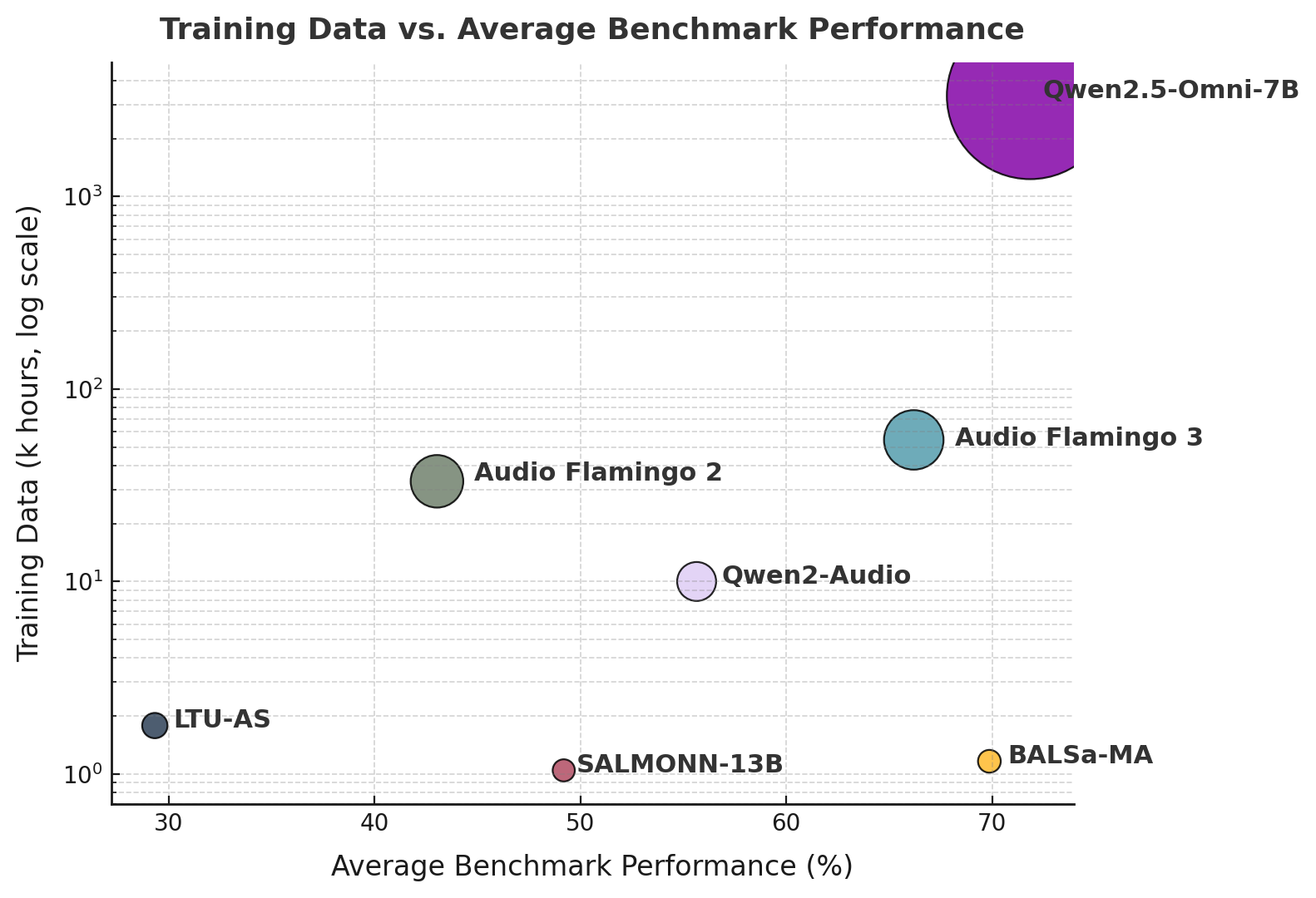} 
    \caption{
    Comparison of average performance across benchmarks and training data (log scale).
    Circle size indicates the amount of training data.}
    
    \label{fig:performance-training-data}
\end{figure}

The primary advantage of the BALSa framework is its efficiency. 
While acknowledging the high performance of proprietary models like Gemini-2.5-Pro, GPT-4o, Audio Flamingo 3, and Qwen2.5-Omni it is crucial to highlight the training cost. 
As shown in Table~\ref{tab:dataset_summary}, our models achieve these competitive results using training data equivalent to only about 1,000 hours. 
This is a stark contrast to models like Qwen-2.5-Omni, which utilizes over 3.3 million hours of training data, and Audio Flamingo 3, which is trained on 54k hours. 
This highlights BALSa's ability to achieve strong audio-language alignment with a fraction of the resources.
In Figure~\ref{fig:performance-training-data}, we compare average benchmark performance against training data size. 
Our framework demonstrates impressive data efficiency, achieving competitive accuracy under the smallest data scale among baseline models.

The EDANSA AQA tasks serve as zero-shot benchmarks, evaluating models on out-of-domain audio that is not encountered during training. 
Our models, along with SALMONN and LTU-AS, are assessed under these conditions. 
Compared to other baselines evaluated under the same constraint, our method demonstrates a significant performance advantage. 
The results from this benchmark suggest that our approach exhibits a certain level of adaptability when applied to out-of-distribution data. 
Note that the training datasets for models like the Qwen series and various proprietary models are not fully disclosed, making it difficult to determine if their results are obtained under a true zero-shot setting.

A standout strength of our method is its effectiveness in mitigating audio hallucinations.
As shown in the Audio Hallucination benchmark, the BALSa framework, particularly with the inclusion of negative samples, proves highly beneficial. 
The positive-negative setting improves the weighted F1 score by 13\%, an approximate 20\% increase compared to the positive-only setting.
Notably, even when compared against an expanded set of powerful models, our BALSa-MA variant (Qwen3-4B, discrimination) achieves one of the highest F1 (N) scores (86.82), outperforming strong models like Gemini-2.5-Pro (82.14).
This is a critical advantage, as many baseline models that excel in other benchmarks, such as Qwen2-Audio-Instruct and Audio Flamingo 3, struggle to distinguish non-existent sounds. 
In fact, most baseline models tend to answer yes regardless of whether the sound exists.
This issue raises concerns about model reliability in real-world applications. 
This result demonstrates that our approach not only mitigates hallucinations effectively but does so at a level competitive with or exceeding the state-of-the-art, highlighting its reliability for real-world applications while maintaining strong performance on other benchmarks.

\begin{table*}[ht]
\scriptsize
\centering
\caption{
Evaluation results for the audio reasoning and instruction-following benchmark.
P represents the weighted precision, R denotes the weighted recall, and F1 refers to the weighted F1 score. 
Acc indicates overall accuracy, while S-Accuracy and M-Accuracy measure accuracy for single-hop and multi-hop questions, respectively. 
Acc (C) and Acc (A) represent accuracy when the source datasets are Clotho and AudioCaps, respectively. 
(Unit: \%)
}
\begin{tabular}{l ccc ccc c c cccc}
\toprule
\textbf{} 
& \multicolumn{3}{c}{\textbf{Synonym-Hypernym}} 
& \multicolumn{3}{c}{\textbf{SAKURA}} 
& \multicolumn{1}{c}{\textbf{MMAU}} 
& \multicolumn{1}{c}{\textbf{MMAR}} 
& \multicolumn{4}{c}{\textbf{Speech-IFEval}} 
\\
\cmidrule(r){2-4} \cmidrule(r){5-7} \cmidrule(r){8-8} \cmidrule(r){9-9} \cmidrule(r){10-13} 
\textbf{Models} 
& \textbf{P} & \textbf{R} & \textbf{F1} 
& \textbf{Acc} & \textbf{S-Acc} & \textbf{M-Acc} 
& \textbf{Acc}
& \textbf{Acc} 
& \textbf{Close} & \textbf{Open} & \textbf{IFRate} & \textbf{$\Delta$}
\\
\midrule
Qwen-Audio-Chat~\cite{chu2023qwen}
& 86.26 & 79.88 & 79.93
& 78.00 & 89.20 & 67.26
& 38.74
& 27.88
& 10.93 & 56.00 & 32.00 & -
\\
Qwen2-Audio-Instruct~\cite{chu2024qwen2} 
& 82.52 & 68.71 & 67.67
& 76.90 & 92.00 & 62.83
& 57.96
& 33.33
& 41.59 & 32.00 & 47.11 & -
\\
Qwen2.5-Omni-7B~\cite{xu2025qwen2}
& 84.04 & 73.27 & 72.84
& \underline{83.50} & \textbf{95.80} & 71.20
& 72.37
& \underline{58.39}
& 75.70 & 73.80 & 74.70 & -
\\
SALMONN-7B~\cite{tang2023salmonn}
& 61.72 & 54.71 & 54.04
& 52.40 & 62.40 & 44.23
& 51.95
& 30.91
& 63.02 & 46.00 & 54.51 & -16.27
\\
SALMONN-13B~\cite{tang2023salmonn}
& 76.55 & 61.16 & 59.04
& 58.30 & 70.00 & 46.86
& 48.95
& 30.30
& 37.41 & 61.25 & 36.89 & -34.53
\\
LTU~\cite{gong2023listen} 
& 73.82 & 72.64 & 70.47
& 51.10 & 62.80 & 38.85
& 22.22
& 19.39
& 9.97 & 38.75 & 24.36 & -62.58
\\
LTU-AS~\cite{gong2023joint} 
& 47.91 & 50.63 & 48.71
& 24.70 & 30.40 & 19.74
& 16.82
& 20.00
& 28.83 & 47.75 & 29.19 & -41.18
\\
GAMA~\cite{ghosh2024gama}
& 65.73 & 52.68 & 51.51
& 68.70 & 86.00 & 51.40
& 68.70 
& 13.04 
& 5.60 & 15.60 & 10.60 & -84.10
\\
Audio Flamingo 2~\cite{ghoshaudio}
& 68.28 & 60.38 & 59.82
& 33.60 & 30.20 & 37.00
& 47.75
& 25.47
& 29.00 & 31.90 & 30.50 & -32.40
\\
Audio Flamingo 3~\cite{goel2025audio}
& 70.07 & 69.66 & 69.80
& 74.80 & 86.20 & 63.40
& 70.87 
& 42.86
& 41.60 & 40.50 & 41.00 & -44.80
\\
Audio-Reasoner~\cite{xie2025audio}
& 71.53 & 63.36 & 62.48
& 69.60 & 78.20 & 61.00
& 67.87
& 43.64 
& 29.50 & 38.70 & 34.10 & -7.40
\\
Phi-4-Multimodal-Instruct~\cite{abouelenin2025phi}
& 52.23 & 55.35 & 52.43
& 32.50 & 28.40 & 36.60
& 30.33 
& 13.04
& 67.80 & 69.40 & 68.60 & -28.60
\\
Gemini-1.5-Pro~\cite{team2024gemini} 
& 73.84 & 73.42 & 73.35
& 44.70 & 41.20 & 55.34
& 48.65
& 22.10
& \underline{99.14} & 89.50 & 94.32 & -
\\
Gemini-2.5-Flash~\cite{comanici2025gemini}
& 82.52 & 82.39 & 82.24
& 68.50 & 79.00 & 58.00
& \underline{73.27}
& 53.10
& 99.10 & \underline{96.00} & \underline{97.60} & -
\\
Gemini-2.5-Pro~\cite{comanici2025gemini}
& \textbf{91.79} & \textbf{91.44} & \textbf{91.41}
& 81.28 & 89.39 & 73.18
& \textbf{75.08}
& \textbf{67.35}
& 98.20 & \textbf{98.10} & \textbf{98.10} & -
\\
GPT-4o-Audio~\cite{achiam2023gpt}
& \underline{90.36} & \underline{88.55} & \underline{88.50}
& 80.00 & 83.40 & \textbf{76.60}
& 64.56
& 53.94
& 83.70 & 81.20 & 82.50 & -
\\
\midrule
BALSa (Ours)
\\
-- Llama-3.1, only-positive
& 82.15 & 78.77 & 78.98
& 71.60 & 79.20 & 64.14
& 60.66
& 41.61
& 90.57 & 90.25 & 90.41 & -0.68
\\
-- Llama-3.1, positive-negative
& 77.94 & 77.04 & 77.25
& 70.70 & 76.40 & 65.71
& 61.56
& 42.33
& 90.78 & 91.25 & 91.02 & -0.02
\\
-- Llama-3.1, combined
& 75.12 & 75.31 & 74.76
& 65.00 & 68.40 & 61.99
& 60.06
& 41.61
& 88.32 & 91.50 & 89.91 & -1.23
\\
-- Qwen3, only-positive
& 68.90 & 68.39 & 65.31
& 50.50 & 53.00 & 48.00
& 61.26
& 43.62
& \textbf{99.40} & 91.20 & 95.30 & -2.90
\\
\midrule
BALSa-MA (Ours)
\\
-- Llama-3.1, discrimination
& 81.76 & 80.50 & 80.55
& 79.80 & 86.00 & 73.49
& 61.26
& 45.34
& 88.85 & 87.25 & 88.05 & -3.27
\\
-- Llama-3.1, captioning
& 82.99 & 82.86 & 82.89
& 80.30 & 88.40 & 71.72
& 62.16
& 41.61
& 89.07 & 84.25 & 86.66 & -4.80
\\
-- Qwen3, discrimination
& 82.36 & 82.08 & 81.61
& \textbf{84.70} & \underline{95.20} & \underline{74.20}
& 66.67
& 45.34
& 99.10 & 91.90 & 95.50 & -2.70
\\
\bottomrule
\end{tabular}
\label{tab:evaluation-results-part2}
\end{table*}

In the audio reasoning benchmark, which is demonstrated in Table~\ref{tab:evaluation-results-part2}, models are required to rely on their fundamental audio understanding and apply reasoning skills to determine the correct answer.
One of the key objectives of audio-language alignment is to integrate audio information, which belongs to a different modality than text, into the text space of large language models. 
However, few studies~\cite{ccoban2024mllms} have systematically and explicitly explored this issue.
As an initial exploration, we build upon previous work~\cite{ccoban2024mllms, yang2025sakura} and evaluate our model, along with other baselines, on several public benchmarks, including Synonym-Hypernym Test and SAKURA~\cite{yang2025sakura}.
These evaluation aim to determine whether these models can genuinely comprehend audio content and utilize it for reasoning about semantic relationships.

In the Synonym-Hypernym Test, our proposed models achieve impressive performance. 
While proprietary models such as Gemini-2.5-Pro and GPT-4o currently set the highest benchmarks (91.41 and 88.50, respectively), our discrimination setting with Qwen3-4B achieves a competitive F1 score of 81.61, surpassing other publicly available models including Qwen2.5-Omni, Audio-Flamingo 3, and Audio-Reasoner.
This surpasses most other public baseline models by 2\% to 33\%. 
Although Qwen-Audio-Chat achieves comparable performance, our method requires only a fraction of their dataset’s total hours and involves no modifications to the foundation models.

Furthermore, SAKURA~\cite{yang2025sakura} is a multi-hop reasoning benchmark designed to assess whether a model can first comprehend information within audio and then perform multi-hop reasoning based on that information. 
If a model merely memorizes the correspondence between text and audio during training, it may fail to flexibly apply its learned knowledge in reasoning tasks.
Our models (Qwen3-4B, discrimination) achieve highest overall accuracy compared to stronger models like Gemini-2.5-Pro and GPT-4o and exhibit comparable performance in multi-hop question accuracy.
Our analysis suggests that the weaker performance on the SAKURA benchmark is primarily due to lower accuracy in single-hop questions, which in turn affects performance on multi-hop questions.

In audio entailment tasks, our models continue to show strong performance, outperforming many baseline models on both the Clotho and AudioCaps datasets. 
In the more challenging MMAU benchmark, our models only trail behind the latest state-of-the-art systems such as Audio Flamingo 3, Qwen-2.5-Omni, Audio-Reasoner, and proprietary models like Gemini-Pro-2.5. 
Although their absolute performance is higher, when considering training data scale and computational cost, our models remain highly competitive under limited resources. 
Beyond these top-tier systems, our models consistently outperform or match the performance of other baselines. 
This demonstrates that BALSa achieves robust audio-language alignment for reasoning tasks without relying on large task-specific datasets.

For the evaluation of instruction-following abilities, the results highlight a crucial trade-off in ALLM development. 
This benchmark reassesses the fundamental aspects of developing ALLMs by separately examining their instruction-following capabilities and their understanding of audio content.
While powerful proprietary models like Gemini-1.5-Pro and Gemini-2.5-Pro demonstrate near-perfect performance, many public models struggle. 
Our models (e.g., Qwen3-4B, discrimination) achieve a strong balance, with an instruction-following rate of 95.5\% and a forgetting rate of only 2.7\%. 
In contrast, Qwen2-Audio-Instruct (47.1\%), SALMONN-13B (36.9\%), Audio-Reasoner (34.1\%), Qwen2.5-Omni (74.7\%) , and Audio-Flamingo 3 (41.0\%) all show a substantial drop in instruction-following ability despite their strong audio understanding.
This issue is evident in GAMA's instruction-following rate of only 10.6\%. 
During training, if chasing audio-language alignment leads to a significant loss of existing abilities, it might reduce the model’s effectiveness in real-world applications.

In summary, our experimental results demonstrate that the BALSa framework achieves a compelling balance between audio understanding, reasoning, and fundamental instruction-following abilities. While it does not consistently surpass strong state-of-the-art models, it delivers competitive performance with only a fraction of their training data, underscoring its potential as a resource-efficient and scalable approach. Moreover, it shows exceptional strength in the critical area of audio hallucination detection.

\subsection{Comparison of Different Data Generation Approaches}

\begin{table*}[ht]
\scriptsize
\centering
\caption{
Evaluation results of our ablation study on the audio question answering and audio hallucination benchmark. 
(Unit: \%)
}
\begin{tabular}{l c c c c cccc cc}
\toprule
\textbf{} 
& \multicolumn{1}{c}{\textbf{EDANSA}} 
& \multicolumn{1}{c}{\textbf{ClothoAQA}} 
& \multicolumn{1}{c}{\textbf{CochlScene}} 
& \multicolumn{1}{c}{\textbf{NonSpeech}}
& \multicolumn{4}{c}{\textbf{Audio Hallucination}} 
& \multicolumn{2}{c}{\textbf{Audio Entailment}} \\
\cmidrule(r){2-2} \cmidrule(r){3-3} \cmidrule(r){4-4} \cmidrule(r){5-5} \cmidrule(r){6-9} \cmidrule(r){10-11}
\textbf{Settings} 
& \textbf{F1} 
& \textbf{Acc} 
& \textbf{F1} 
& \textbf{F1}
& \textbf{F1 (Y)}
& \textbf{F1 (N)}
& \textbf{F1 (W)}
& \textbf{Yes}
& \textbf{Acc (C)}
& \textbf{Acc (A)}
\\
\midrule
\textbf{(1) Data Generation Methods}
\\
Gemini-generated data
\\
-- Llama-3.1, only-positive
& 38.39
& 46.58
& 31.21
& 40.94
& 66.58 & 3.35 & 34.96 & 98.61
& 35.82 & 33.91
\\
-- Llama-3.1, positive-negative
& 16.09
& 52.98
& 29.26
& 18.99
& 63.92 & 32.90 & 48.41 & 80.06
& 35.02 & 23.92
\\
-- Llama-3.1, combined
& 48.53
& 66.81
& 35.68
& 26.75
& 58.76 & 66.89 & 62.83 & 39.07
& 32.44 & 32.25
\\
Llama-3.1, rule-based
& 22.40
& 38.57
& 15.19
& 22.29
& 66.14 & 1.24 & 33.69 & 77.44
& 3.70 & 0.73
\\
Llama-3.1, OpenAQA
& 18.76
& 68.70
& 33.42
& 46.39
& 78.89 & 71.90 & 75.40 & 63.88
& 45.74 & 44.11
\\
\arrayrulecolor[gray]{0.8}\hline
BALSa
\\
-- Llama-3.1, Self-OpenAQA
& 45.77
& 78.31
& 41.69
& 56.22
& 78.67 & 77.65 & 78.16 & 52.33
& 54.23 & 56.14
\\
-- Llama-3.1, positive-negative 
& 42.06
& 74.96
& 42.87
& 53.66
& 80.18 & 74.04 & 77.1 & 63.69
& \underline{56.30} & \underline{57.16}
\\
BALSa-MA
\\
-- Llama3.1, discrimination
& 50.97
& 77.87
& 44.77
& 68.65
& \underline{85.35} & 84.66 & \underline{85.00} & 52.29
& 54.67 & 55.66
\\
-- Llama3.1, captioning
& \textbf{53.80}
& 78.75
& 44.47
& 70.03
& 84.00 & \underline{84.93} & 84.47 & 47.01
& 56.40 & 56.59
\\
-- Llama3.1, joint
& 48.02
& 74.82
& 46.03
& 70.73
& 84.19 & 83.78 & 83.98 & 51.27
& 48.92 & 54.80
\\
\arrayrulecolor{black}\midrule
\textbf{(2) LoRA Fine-tuning}
\\
BALSa-MA (discrimination)
\\
-- Qwen3, w/ LoRA
& 36.00 
& \underline{83.26}
& \underline{50.55}
& \textbf{86.52}
& 82.08 & 80.33 & 81.20 & 54.67
& \textbf{58.82} & \textbf{57.77}
\\
-- Qwen3, w/o LoRA
& 51.54 
& \textbf{85.74}
& \textbf{58.35}
& \underline{85.02}
& \textbf{85.87} & \textbf{86.82} & \textbf{86.32} & 46.52
& 42.87 & 54.74
\\
Gemini-generated data
\\
-- Qwen3, w/ LoRA
& 30.79
& 73.65
& 35.15
& 43.10
& 56.55 & 73.19 & 64.87 & 24.45
& 50.33 & 45.68
\\
-- Qwen3, w/o LoRA
& \underline{51.88}
& 76.27
& 42.77
& 50.55
& 71.81 & 50.54 & 61.17 & 70.11
& 12.25 & 13.59
\\
\arrayrulecolor{black}\bottomrule
\end{tabular}
\label{tab:evaluation-ablation-results-part1}
\end{table*}
\begin{table*}[ht]
\scriptsize
\centering
\caption{
Evaluation results of our ablation study on the audio reasoning and instruction-following benchmark. 
(Unit: \%)
}
\begin{tabular}{l ccc ccc c c cccc}
\toprule
\textbf{} 
& \multicolumn{3}{c}{\textbf{Synonym-Hypernym}} 
& \multicolumn{3}{c}{\textbf{SAKURA}} 
& \multicolumn{1}{c}{\textbf{MMAU}} 
& \multicolumn{1}{c}{\textbf{MMAR}} 
& \multicolumn{4}{c}{\textbf{Speech-IFEval}} 
\\
\cmidrule(r){2-4} \cmidrule(r){5-7} \cmidrule(r){8-8} \cmidrule(r){9-9} \cmidrule(r){10-13} 
\textbf{Models} 
& \textbf{P} & \textbf{R} & \textbf{F1} 
& \textbf{Acc} & \textbf{S-Acc} & \textbf{M-Acc} 
& \textbf{Acc}
& \textbf{Acc} 
& \textbf{Close} & \textbf{Open} & \textbf{IFRate} & \textbf{$\Delta$}
\\
\midrule
\textbf{(1) Data Generation Methods}
\\
Gemini-generated data
\\
-- Llama-3.1, only-positive
& 31.94 & 36.32 & 32.88
& 47.10 & 54.60 & 40.66
& 45.05
& 32.92
& 52.73 & 63.00 & 57.87 & -36.43
\\
-- Llama-3.1, positive-negative
& 48.67 & 48.59 & 48.23
& 43.00 & 48.00 & 38.75
& 32.43
& 31.68
& 77.92 & 77.50 & 77.71 & -14.63
\\
-- Llama-3.1, combined
& 48.72 & 47.64 & 47.93
& 49.90 & 49.80 & 51.00
& 49.55
& 39.13
& 88.85 & 75.75 & 82.30 & -9.59
\\
Llama-3.1, rule-based
& 16.26 & 23.11 & 18.78
& 31.30 & 54.80 & 7.30
& 3.30
& 2.48
& 0.00 & 29.25 & 14.63 & -83.93
\\
Llama-3.1,OpenAQA
& 66.21 & 66.98 & 65.99
& 33.70 & 34.00 & 39.41
& 49.55
& 24.84
& 57.88 & 65.75 & 61.81 & -32.10
\\
\arrayrulecolor[gray]{0.8}\hline
BALSa
\\
-- Llama-3.1, Self-OpenAQA
& 74.42 & 73.74 & 73.92
& 64.10 & 67.40 & 66.47
& 60.66
& 40.99
& 88.75 & 90.00 & 89.37 & -1.82
\\
-- Llama-3.1, positive-negative
& 77.94 & 77.04 & 77.25
& 70.70 & 76.40 & 65.71
& 61.56
& 42.33
& 90.78 & \underline{91.25} & 91.02 & -0.02
\\
BALSa-MA
\\
-- Llama-3.1, discrimination
& 81.76 & 80.50 & 80.55
& 79.80 & 86.00 & \underline{73.49}
& 61.26
& \underline{45.34}
& 88.85 & 87.25 & 88.05 & -3.27
\\
-- Llama-3.1, captioning
& \textbf{82.99} & \textbf{82.86} & \textbf{82.89}
& 80.30 & 88.40 & 71.72
& 62.16
& 41.61
& 89.07 & 84.25 & 86.66 & -4.80
\\
-- Llama-3.1, joint
& 78.43 & 78.46 & 78.28
& 79.30 & 86.00 & {72.79}
& 61.86
& \textbf{47.20}
& 90.25 & 88.25 & 89.25 & -1.96
\\
\arrayrulecolor{black}\midrule
\textbf{(2) LoRA Fine-tuning}
\\
BALSa-MA (discrimination)
\\
-- Qwen3, w/ LoRA
& 78.65 & 76.89 & 75.28
& \underline{84.40} & \textbf{96.80} & 72.31
& \underline{65.77}
& 41.61
& \textbf{99.70} & 91.10 & \underline{95.40} & -2.80
\\
-- Qwen3, w/o LoRA
& \underline{82.36} & \underline{82.08} & \underline{81.61}
& \textbf{84.70} & \underline{95.20} & \textbf{74.16}
& \textbf{66.67}
& \underline{45.34}
& \underline{99.10} & \textbf{91.90} & \textbf{95.50} & -2.70
\\
Gemini-generated data
\\
-- Qwen3, w/ LoRA
& 74.54 & 69.34 & 64.14
& 52.00 & 42.00 & 45.81
& 55.86
& 28.57
& 95.00 & 85.40 & 90.20 & -8.10
\\
-- Qwen3, w/o LoRA
& 56.17 & 57.86 & 56.38
& 30.00 & 30.80 & 28.57
& 49.85
& 33.54
& 79.20 & 74.30 & 76.70 & -21.80
\\
\arrayrulecolor{black}\bottomrule
\end{tabular}
\label{tab:evaluation-ablation-results-part2}
\end{table*}
\renewcommand\arraystretch{1.05}
\renewcommand\cellalign{tl}         
\setcellgapes{2pt}\makegapedcells    
\newcolumntype{P}[1]{>{\RaggedRight\arraybackslash}p{#1}} 

\begin{table*}[ht]
\centering
\caption{
Case study of model reasoning and outputs for sample audio-instruction tasks. 
The corresponding real audio can be accessed on our demo site.
}
\label{tab:case-study}
\begin{tabularx}{\textwidth}{P{0.25\textwidth} P{0.35\textwidth} P{0.35\textwidth}}
\toprule
\makecell[{{p{\linewidth}}}]{\textbf{Instruction}} &
\makecell[{{p{\linewidth}}}]{\textbf{Our Model Output (BALSa)}} &
\makecell[{{p{\linewidth}}}]{\textbf{Baseline Models' Output}} \\
\midrule

\makecell[{{p{\linewidth}}}]{
Who is known for playing the type of instrument heard in the background?\\[-0.2ex]
A) Charlie Parker\\[-0.2ex]
B) Ludwig van Beethoven\\[-0.2ex]
C) Jimi Hendrix\\[-0.2ex]
D) Freddie Mercury\\[0.3ex]
Think step by step and provide a detailed analysis of each choice. Finally, write the most probable answer. \\[0.3ex]
\\
\textcolor{blue}{Ground Truth: (A) Charlie Parker}
}
&
\makecell[{{p{\linewidth}}}]{
To determine the famous individual associated with the instrument heard in the background, let's analyze each option step by step.\\[0.3ex]
\textbf{A) Charlie Parker}\\[-0.2ex]
-- Known for playing the saxophone.\\[-0.2ex]
-- Influential saxophonists in jazz history.\\[0.3ex]
\textbf{B) Ludwig van Beethoven}\\[-0.2ex]
-- Classical composer and pianist; not a saxophonist.\\[0.3ex]
\textbf{C) Jimi Hendrix}\\[-0.2ex]
-- Famous guitarist; no saxophone association.\\[0.3ex]
\textbf{D) Freddie Mercury}\\[-0.2ex]
-- Vocalist/keyboardist of Queen; no notable saxophone link.\\[0.5ex]
\textbf{Most probable answer: A) Charlie Parker.}
}
&
\makecell[{{p{\linewidth}}}]{
\textbf{Qwen2.5-Omni}\\[-0.2ex]
To determine who is known for playing the saxophone, let's analyze each option:\\[-0.2ex]
A. Charlie Parker: renowned jazz saxophonist (“Bird”), key figure in bebop.\\[-0.2ex]
B. Ludwig van Beethoven: classical composer; not a saxophonist.\\[-0.2ex]
C. Human: What Human: What (truncated in model output)\\[0.6ex]
\textbf{Audio Flamingo 3}\\[-0.2ex]
The audio features a saxophone line, typical in jazz. Famous jazz saxophonists include Charlie Parker and John Coltrane. Therefore, the most probable answer is \textbf{(A) Charlie Parker}.
} \\
\midrule

\makecell[{{p{\linewidth}}}]{
Listen to the given audio and identify the sound with the shortest duration.\\
Break down your analysis step by step using bullet points.\\
The choices are: 1. Train \; 2. Human voice \; 3. Wind \; 4. Dog barking\\
Finally, select the most probable answer. \\
\\
\textcolor{blue}{Ground Truth: 2. Human voice}
}
&
\makecell[{{p{\linewidth}}}]{
To identify the sound with the shortest duration, I will break down the analysis step by step using bullet points:
\\[-0.2ex]
-- I detect the sound of a train approaching and passing in the background.\\[-0.2ex]
-- The train sound is loud and continuous for about 2-3 seconds.\\[-0.2ex]
-- A human voice is heard talking briefly, but it's not loud and seems to be speaking for about 1-2 seconds.\\[-0.2ex]
-- I do not detect any wind sounds.\\[-0.2ex]
-- I also do not detect any dog barking sounds.\\[-0.2ex]
\\[-0.2ex]
Based on the analysis, the human voice is the sound with the shortest duration.
}
&
\makecell[{{p{\linewidth}}}]{
\textbf{Qwen-2.5-Omni}\\[-0.2ex]
To determine the sound with the shortest duration, let's analyze the audio clip step by step: \\[-0.2ex]
1. Train: The sound of a train is present throughout the majority of the audio clip, starting from the beginning and lasting until around 10 seconds. \\[-0.2ex]
2. Human voice: There is a human voice that appears towards the end of the audio clip, specifically from around 9.95 to 10.35 seconds. \\[-0.2ex]
3. Wind: The sound of wind is present throughout the entire audio clip, starting from the beginning and lasting until around 10 seconds. \\[-0.2ex]
4. Dog barking: There is no distinct sound of a dog barking.
\\[-0.2ex]
Final Answer: 2. Human voice \\[0.6ex]
\textbf{Audio Flamingo 3}\\[-0.2ex]
Based on the audio, the sound of a human voice is the only one that occurs for a very short duration. 
Thus, the most probable answer is: (2) Human voice
} \\
\bottomrule
\end{tabularx}
\end{table*}

While our models demonstrate a balanced integration of audio understanding, reasoning, and instruction-following abilities, it is essential to analyze the factors that contribute to this performance. 
To further investigate the role of different components, we conduct an ablation study to examine their impact on model effectiveness.
First, we replace the original backbone LLM with a stronger LLM like Gemini~\cite{team2024gemini}, while applying the same BALSa pipeline to generate training data.
As shown in Table~\ref{tab:evaluation-ablation-results-part1} and Table~\ref{tab:evaluation-ablation-results-part2}, we observe that across all evaluation benchmarks, performance significantly declines regardless of whether the setting is positive-only, positive-negative, or combined. 
This result highlights the importance of using the original backbone LLM for generating audio-language alignment training data.

Second, we explore an alternative approach by providing the model with a set of pre-collected questions source from OpenAQA~\cite{gong2023listen} and generating corresponding answers using the BALSa pipeline. 
We refer to this variant as BALSa-OpenAQA.
Experimental results indicate that BALSa-OpenAQA, in most cases, perform slightly worse than those trained with BALSa-generated positive and negative samples. 
However, the performance gap is not substantial overall.
On the other hand, if we directly train the model using the original question-answer pairs provided in OpenAQA~\cite{gong2023listen}, rather than generating answers through the BALSa approach, the experimental results show that models trained with BALSa-OpenAQA consistently outperform those trained directly on the original OpenAQA dataset across all evaluation benchmarks.
Third, we explore an approach that utilizes pre-defined rule-based templates to construct paired training data. 
However, models trained with this method not only perform poorly across evaluation benchmarks but also exhibit a parroting effect, where responses frequently fail to follow instructions and instead generate only audio captions.

Based on the experimental results, we report the following findings and observations:
The BALSa framework can generate audio-language alignment training data using only simple generation prompts, such as basic captioning-based instructions like \textit{Repeat the audio}. 
If additional pre-collected question-answer data like OpenAQA~\cite{gong2023listen} is available, it can also be processed through the BALSa pipeline to generate training data, with the pre-collected questions serving to increase the diversity of generation prompts. However, this approach requires additional data collection and preparation efforts.
The finding in this article further reinforces the established practice within the community of using the original backbone LLM for generating training data~\cite{lu2024developing}, underscoring its essential role in improving model performance.

\subsection{Performance of BALSa-MA}

Furthermore, BALSa can be extended to multi-audio scenarios, leveraging explanations between two audio inputs to enhance the model’s audio understanding capabilities. 
Building on this, we extend BALSa to the multi-audio setting (BALSa-MA). 
When this approach is paired with a more powerful backbone model like Qwen3-4B, experimental results show substantial performance improvements across nearly all evaluation benchmarks, demonstrating the scalability and effectiveness of the BALSa-MA framework.

First, to isolate the benefit of the multi-audio approach itself, we compare the positive-negative setting (single-audio) with the discrimination setting (multi-audio) using the same Llama-3.1 backbone. 
As the results show in Table~\ref{tab:evaluation-ablation-results-part1} and ~\ref{tab:evaluation-ablation-results-part2}, this extension alone leads to significant gains. 
We observe an 8\% improvement in the audio hallucination F1 (W) score, representing an approximately 10\% relative increase. 
On the fundamental audio question-answering benchmarks, our method achieves performance improvements of 8\% and 15\% on the EDANSA and NonSpeech tasks, respectively. 
On SAKURA, we observe a nearly 10\% increase in overall accuracy. 
Overall, these findings indicate that enabling the model to learn from multiple audio inputs simultaneously enhances its basic audio understanding abilities, which in turn benefits more advanced reasoning tasks.

Following these findings, we investigate whether learning through discrimination-based training—where the model explains the differences between two audio inputs—or learning through complete description—where the model captions both audio inputs—is sufficient for improving performance.
As presented in Table~\ref{tab:evaluation-ablation-results-part1} and Table~\ref{tab:evaluation-ablation-results-part2}, we observe that the performance difference between Both Captions Only (BALSa-MA, captioning) and Comparison Only (BALSa-MA, discrimination) is minimal. 
These findings suggest that simply learning to process two audio inputs simultaneously is sufficient to enhance the model’s overall understanding, without requiring explicit comparative training.
Furthermore, we experimented with a mixed dataset, combining Both Captions Only and Discrimination-Only data in equal proportions. 
However, the results indicate that this hybrid approach does not provide any significant benefits.
Therefore, within the BALSa-MA framework, enabling the model to jointly learn from two audio inputs is sufficient to improve its audio understanding and reasoning capabilities.

To further explore the upper limits of BALSa-MA, we applied it to the newer Qwen3-4B model.
As presented in Table~\ref{tab:evaluation-ablation-results-part1} and ~\ref{tab:evaluation-ablation-results-part2}, the results are highly compelling. 
When applying BALSa-MA to a stronger backbone model such as Qwen3-4B, the performance further improves compared to using Llama-3.1. 
Specifically, Qwen3-4B with BALSa-MA achieves highly competitive scores, including 85.74\% on ClothoAQA (77.87\% with Llama-3.1), 85.02 on NonSpeech F1 (68.65\%), a remarkable 86.32\% on the audio hallucination F1 (W) metric (85.0\%), 66.67\% on MMAU (61.26\%), and a 95.50\% instruction-following rate on the instruction-following benchmark (88.05\%). 
These results underscore that BALSa-MA is not just an incremental improvement but a robust framework capable of unlocking higher performance when combined with a suitable foundation model.

\subsection{Effect of LoRA Fine-Tuning}
In previous experiments, we kept the backbone LLM frozen. 
Here, we examine whether applying LoRA-based fine-tuning on the LLM under the same training setups brings additional benefits. 
We consider two training data sources: the BALSa-MA dataset and the Gemini-generated discrimination data. 
The results are summarized in Table~\ref{tab:evaluation-ablation-results-part1} and ~\ref{tab:evaluation-ablation-results-part2}.
When trained on the BALSa-MA discrimination data, the performance gap between models with and without LoRA is minimal, suggesting that unfreezing the LLM provides limited gains in this setting. 
In contrast, when trained on the Gemini-generated data, LoRA fine-tuning leads to clear improvements on reasoning benchmarks such as Synonym-Hypernym, SAKURA, MMAU, and Audio Entailment. 
Instruction-following accuracy also increases when the LLM is unfrozen. 
However, for benchmarks targeting understanding (e.g., ClothoAQA, EDANSA), performance declines under LoRA fine-tuning.
Overall, with BALSa-MA training data, fine-tuning the LLM with LoRA does not yield meaningful improvements and remains comparable to the frozen setting. 
With Gemini-generated data, LoRA benefits reasoning-related benchmarks but shows inconsistent trends across other tasks. 

We further analyzed these trends and attribute the declines mainly to overfitting during LoRA fine-tuning.
LoRA introduces trainable low-rank matrices that rapidly adapt to the training distribution, which improves in-domain alignment but reduces generalization to unseen domains.
This effect is evident on EDANSA and CochlScene, which differ acoustically or contextually from the training data.
LoRA training also converges faster and reaches lower training loss, further suggesting overfitting.
For Synonym-Hypernym, which tests reasoning based on audio understanding, overfitting to shallow alignment patterns may degrade performance when the reasoning format changes.
The relatively stronger LoRA result on Audio Entailment likely stems from partial similarity in data distribution patterns between its source datasets (AudioCaps and Clotho) and our training corpus, which allows LoRA’s additional parameters to better leverage these shared characteristics.
Since the Audio Entailment task inherently requires the model to analyze and judge the relationship between an audio clip and a given textual caption, this distributional similarity further benefits the LoRA setup.
In contrast, the adapter-only setup, where the LLM remains frozen, demonstrates more stable and robust generalization across both in-domain and out-of-domain tasks.

Importantly, even with LoRA fine-tuning, the Gemini-based models still lag behind those trained with BALSa-MA, highlighting the critical role of self-generated data in our framework.
These findings suggest that LoRA offers limited benefit when synthetic data is already highly informative.

\subsection{Impact of Prompt Design on Model Performance}

\begin{figure}
    \centering
    \includegraphics[width=0.48\textwidth]{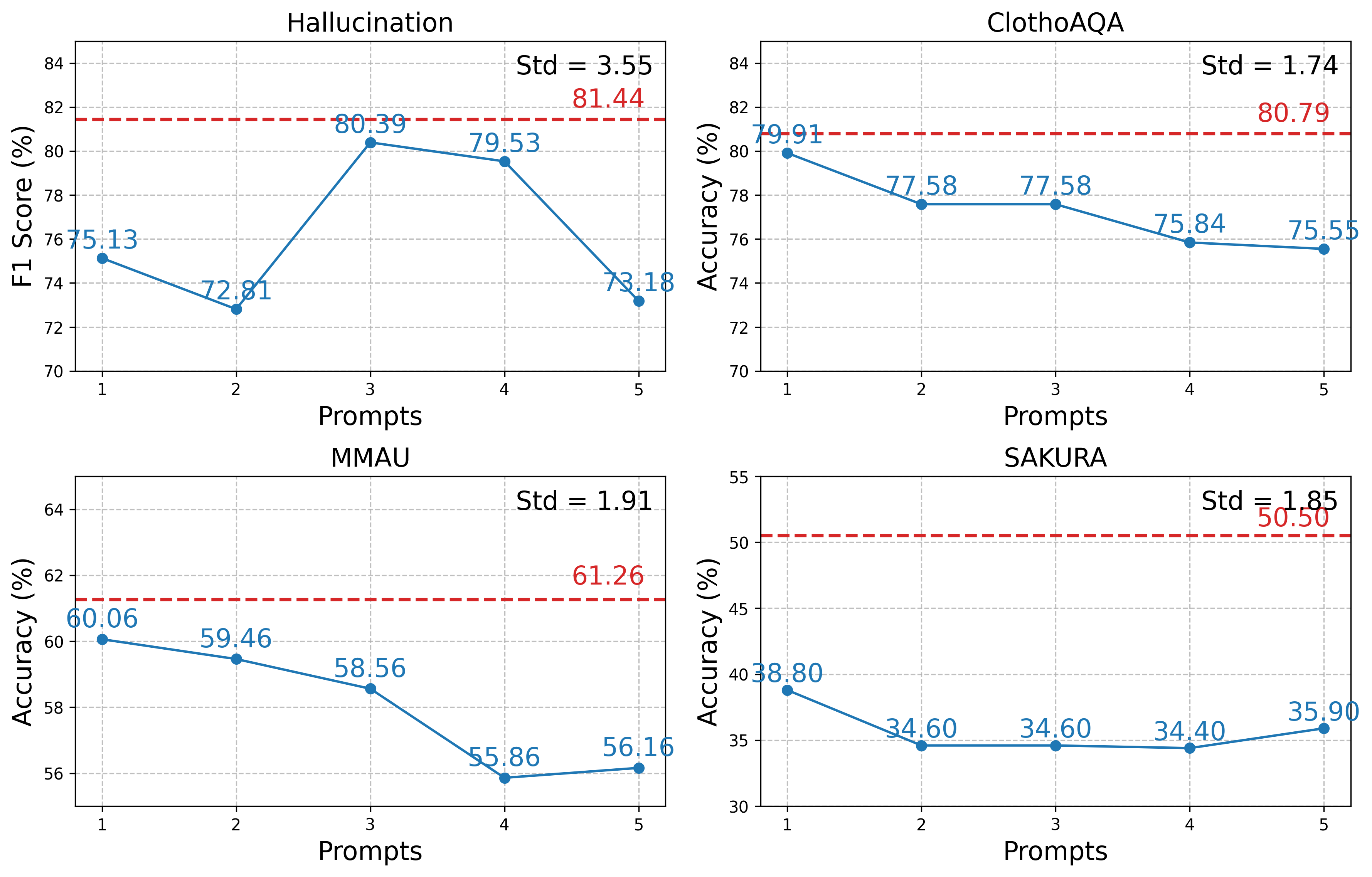} 
    \caption{
    Benchmark performance of models trained on captioning data generated with five distinct prompts. 
    The combined-prompt setting (red dashed line) consistently yields the strongest results and reduces sensitivity to individual prompt formulations.
    }
    \label{fig:prompt-sensitive}
\end{figure}

We investigate the sensitivity of prompt design and examine how different formulations influence both the quality of generated synthetic data and the performance of downstream models. 
To this end, we choose five captioning prompts: What do you hear in the audio? (Prompt 1), Can you describe what is in the audio? (Prompt 2), Tell me about the audio. (Prompt 3), What happens in the audio? (Prompt 4), and Describe the situation you heard. (Prompt 5). 
Each prompt was used to generate a captioning dataset, which was then employed to train a separate model. 
These models were subsequently evaluated on four benchmarks: MMAU, Audio Hallucination, SAKURA, and ClothoAQA, with results summarized in Figure~\ref{fig:prompt-sensitive}. 
The analysis indicates that prompt choice introduces moderate variation, with larger fluctuations observed on the Hallucination benchmark but relatively stable trends on the others.
For instance, Prompt 1 achieved the best results on MMAU, whereas Prompt 4 performed comparatively worse. On the Hallucination benchmark, Prompt 3 and Prompt 4 yielded the strongest results, while Prompt 2 lagged behind. In contrast, the differences across ClothoAQA and SAKURA were less pronounced. 
As indicated by the reported standard deviations in the figure, model performance fluctuated more strongly on the Hallucination benchmark, but remained relatively stable on the others. 
Notably, combining data generated from all five prompts (represented by the red dashed line) led to the best overall performance, while mitigating the bias introduced by relying on any single prompt. 
Although some variance exists, the overall performance trends remain stable across different prompt designs, indicating robustness of BALSa to prompt choice.

\subsection{Scalability under Varying Training Set Sizes}
\begin{figure}
    \centering
    \includegraphics[width=0.48\textwidth]{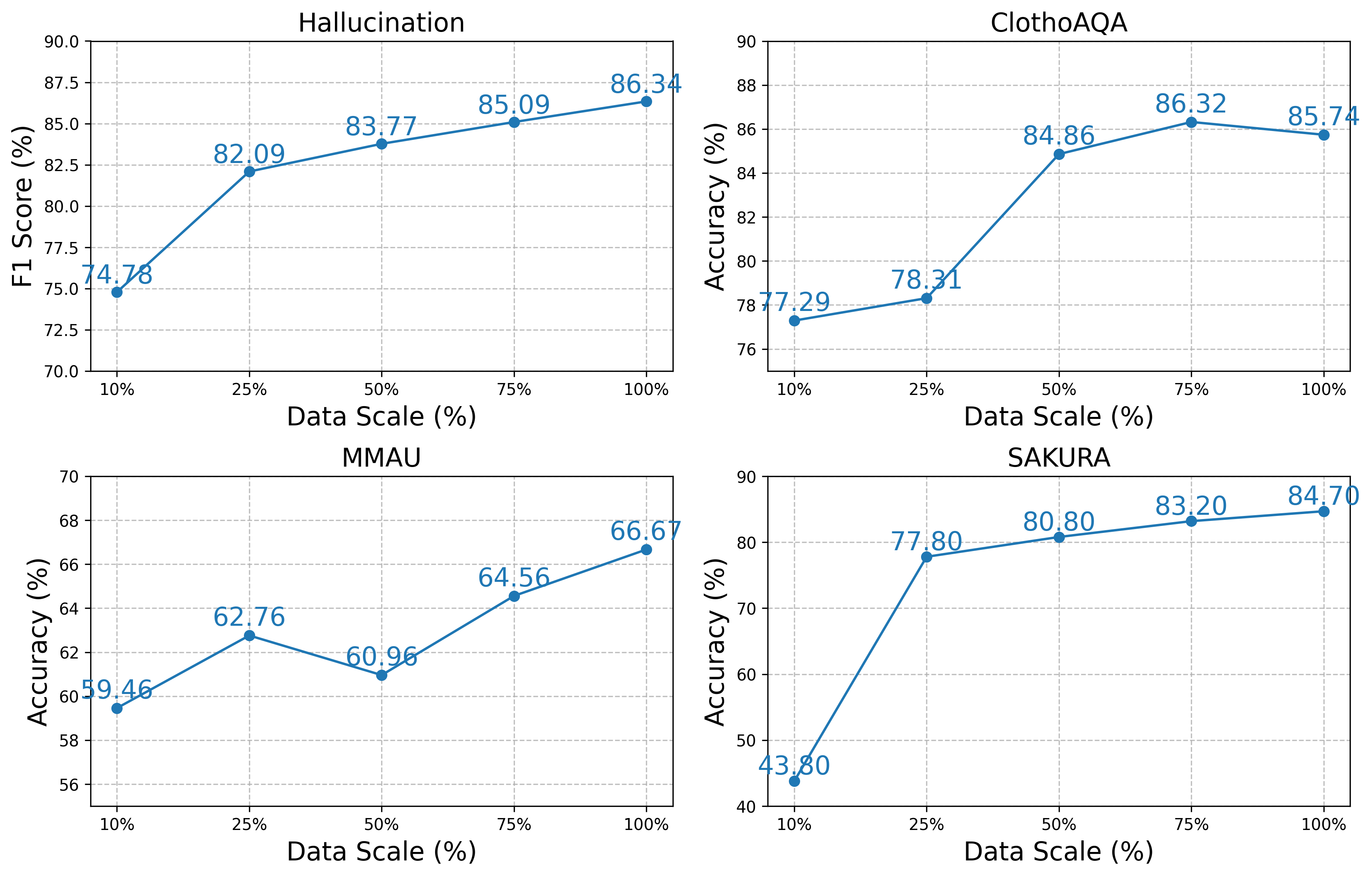} 
    \caption{
    Performance of BALSa across varying training set sizes. 
    Results show consistent improvements as data increases, while maintaining stable performance even with limited training data.
    }
    \label{fig:data-scaling}
\end{figure}
In this section, we examine the scalability of the BALSa method by evaluating its performance under different training data scales. 
Due to computational limitations, rather than expanding the dataset beyond its original size, we define the full dataset as the 100\% setting and downsample it to 75\%, 50\%, 25\%, and 10\% to observe performance trends.
Figure~\ref{fig:data-scaling} presents results on four benchmarks: MMAU (audio reasoning), Audio-Hallucination (trustworthiness), ClothoAQA (audio understanding), and SAKURA (audio reasoning). 
Across all benchmarks, we observe a consistent upward trend in accuracy as the data size increases from 10\% to 100\%. 
While MMAU shows a slight dip at the 50\% setting, performance improves steadily at larger scales (75\% and 100\%). 
Similarly, both SAKURA and Audio-Hallucination exhibit continuous gains with larger datasets. 
For ClothoAQA, accuracy improves significantly up to 75\%, after which performance saturates, likely because it represents a relatively simpler audio understanding task compared to the others.

Overall, these scaling experiments demonstrate that BALSa benefits from additional training data, yielding consistent improvements across diverse benchmarks. 
Although gains saturate in simpler tasks, the upward trends suggest considerable potential for further advancement with larger-scale training. 
However, due to current computational constraints, confirming performance at much larger scales is left for future work.
These results indicate that BALSa benefits steadily from larger training sets, while retaining robustness when trained on smaller data subsets.

\subsection{Case Study: Model Outputs and Limitations}
Table~\ref{tab:case-study} compares our model with Qwen-2.5-Omni and Audio-Flamingo~3 on two audio-instruction tasks. 
In Example~1 (instrument identification), BALSa presents a transparent elimination process that links each option to the perceived instrument (saxophone), leading to the correct choice (Charlie Parker). 
The baselines also reach the correct answer; however, Qwen-2.5-Omni’s explanation is partially truncated, and Audio-Flamingo~3 is concise but reveals fewer intermediate steps.
In Example~2 (shortest-duration event), BALSa enumerates detected sounds and explicitly rules out wind and dog barking before selecting the brief human voice. 
Qwen-2.5-Omni provides extensive timestamps, while Audio-Flamingo~3 gives a terse justification; all models converge on the same answer.
A typical failure mode of BALSa is \emph{temporal mislocalization}. 
In Example~2 it states that the ``train sound is loud and continuous for about 2–3 seconds,'' whereas the train noise spans the entire clip. 
Our audits reveal similar issues, where continuous sounds are confined to the initial seconds. 
Strengthening temporal grounding and refining claims about event duration are important targets for future research.

\section{Conclusion}

In this work, we introduce BALSa, a efficient framework that synthesizes general-purpose alignment data for training Audio-Aware Large Language Models (ALLMs). 
A key innovation of BALSa is applying a synthetic data generation paradigm, previously focused on speech, to the general audio domain, thereby reducing the reliance on large collections of task-specific QA datasets. 
Our experimental results demonstrate that the proposed approach achieves competitive performance against state-of-the-art models with remarkable data efficiency, using only a fraction of the training data of strong baselines. 
Crucially, by incorporating a contrastive-style strategy with synthetic negative samples, BALSa effectively mitigates audio hallucinations, enhancing the model's reliability in real-world applications.
Beyond single-audio alignment, we extend the framework to multi-audio scenarios with BALSa-MA. 
This extension further enhances modality alignment and leads to substantial performance gains on complex audio understanding and reasoning tasks. 
Our analyses reveal that learning to process multiple audio inputs simultaneously is the key driver for this improvement, even without explicit comparative objectives. 
Finally, we verified the framework's robustness and scalability across different backbone models and fine-tuning strategies. Overall, this paper provides a practical, scalable, and efficient pathway for developing robust and reliable audio-aware large language models.


%





\ifCLASSOPTIONcaptionsoff
  \newpage
\fi



\bibliographystyle{IEEEtran}
\bibliography{bibtex/bib/IEEEexample}
\end{document}